\documentclass[fleqn]{aa} 
\usepackage{float}
\usepackage{capt-of}
\usepackage{multicol}
\usepackage[T1]{fontenc}

\DeclareRobustCommand{\VAN}[3]{#2}
\let\VANthebibliography\thebibliography
\def\thebibliography{\DeclareRobustCommand{\VAN}[3]{##3}\VANthebibliography}

\usepackage{graphicx}
\usepackage{url}
\usepackage{xcolor}

\usepackage{amssymb}
\usepackage{placeins}
\usepackage{tabularx}
\usepackage[inline]{enumitem}
\usepackage{tabulary}
\usepackage{booktabs}
\usepackage{multirow}
\usepackage{siunitx}
\usepackage{listings}
\usepackage{lastpage}
\usepackage{orcidlink}
\usepackage[normalem]{ulem}
\DeclareUnicodeCharacter{2248}{$\approx$}
\DeclareUnicodeCharacter{0306}{$\check{c}$}
\usepackage{txfonts}

\usepackage{amsmath}
\usepackage{xspace}
\usepackage{tikz}
\usetikzlibrary{positioning}
\newcommand\NII{[\ion{N}{II}]\xspace} 
\newcommand\OIII{[\ion{O}{III}]\xspace} 
\makeatletter 
  \patchcmd{\NAT@citex}
    {\@citea\NAT@hyper@{%
      \NAT@nmfmt{\NAT@nm}%
      \hyper@natlinkbreak{\NAT@aysep\NAT@spacechar}{\@citeb\@extra@b@citeb}%
      \NAT@date}}
    {\@citea\NAT@nmfmt{\NAT@nm}%
    \NAT@aysep\NAT@spacechar\NAT@hyper@{\NAT@date}}{}{}

  \patchcmd{\NAT@citex}
    {\@citea\NAT@hyper@{%
      \NAT@nmfmt{\NAT@nm}%
      \hyper@natlinkbreak{\NAT@spacechar\NAT@@open\if*#1*\else#1\NAT@spacechar\fi}%
        {\@citeb\@extra@b@citeb}%
      \NAT@date}}
    {\@citea\NAT@nmfmt{\NAT@nm}%
    \NAT@spacechar\NAT@@open\if*#1*\else#1\NAT@spacechar\fi\NAT@hyper@{\NAT@date}}
    {}{}
\makeatother

\defcitealias{2023MNRAS.526.3871K}{Paper I}
\defcitealias{2024A&A...692A..79K}{Paper II}

\titlerunning{\texttt{TODDLERS\,2.0}}
\title{\texttt{TODDLERS\,2.0}: Stellar feedback and observables across diverse IMFs, binary populations, and cloud environments}
\author{
Anand Utsav Kapoor\inst{1}\orcidlink{0000-0002-5187-1725}\thanks{E-mail: anandutsavkapoor@gmail.com}
\and
Andrea Gebek\inst{1}\orcidlink{0000-0002-0206-8231}
\and
Maarten Baes\inst{1}\orcidlink{0000-0002-3930-2757}
\and
Sven De Rijcke\inst{1}\orcidlink{0000-0001-7680-2059}
\and
Arjen van der Wel\inst{1}\orcidlink{0000-0002-5027-0135}
\and
S\'ebastien Vicens-Mouret\inst{2,3}\orcidlink{0009-0000-0151-938X}
\and
Carmelle Robert\inst{2,3}\orcidlink{0000-0003-2344-6593}
}

\institute{
Department of Physics and Astronomy, Proeftuinstraat 86 N3, B-9000 Ghent, Belgium
\and
D\'epartement de Physique, de G\'enie Physique et d'Optique, Universit\'e Laval, Qu\'ebec, QC G1V 0A6, Canada
\and
Centre de Recherche en Astrophysique du Qu\'ebec (CRAQ), Qu\'ebec, QC G1V 0A6, Canada}

\date{Accepted XXX. Received YYY; in original form ZZZ}

\begin{document}
\label{firstpage}

\date{Received [date]; accepted [date]}

\abstract
  {Modeling the feedback-driven evolution of star-forming regions and their multi-wavelength emission is central to interpreting observations of galaxies across cosmic time. The \texttt{TODDLERS} framework couples 1D shell dynamics with \texttt{Cloudy} photoionization to produce UV-to-mm observables. The original framework, however, assumed instantaneous star formation, uniform cloud density, a fixed IMF with single-star evolution, and fixed dust grain properties.}
  {We present \texttt{TODDLERS\,2.0}, which removes these restrictions and extends the framework to cover a broader range of stellar populations, birth-cloud physics, and dust properties.}
  {For stellar feedback and input spectra, we integrate \texttt{pySTARBURST99} (arbitrary IMFs, upper mass limits up to $500\,M_{\odot}$) and \texttt{BPASS} (binary evolution, upper mass limits up to $300\,M_{\odot}$), alongside stochastic IMF sampling for low-mass clusters ($M_* \lesssim 10^4\,M_{\odot}$) and a constant star formation rate mode. The 1D evolution includes non-uniform cloud density profiles and dynamic cloud density evolution driven by escaping ionizing radiation. The \texttt{Cloudy} post-processing is extended with modified grain size distributions and diffuse ionized gas.}
  {Cloud density profile and star formation mode jointly control the fragmentation timescale (the onset of momentum-driven expansion) and shell extent: centrally concentrated profiles fragment earlier, and constant star formation delays fragmentation relative to instantaneous bursts. A top-heavy IMF drives stronger feedback and earlier fragmentation than a standard Kroupa IMF. Dynamic cloud density evolution adds a feedback channel that is most consequential at low metallicity, where the unswept cloud density can drop by three orders of magnitude. At low cluster masses, stochastic sampling introduces order-of-magnitude variance in the feedback, confirming the breakdown of the fully-sampled IMF assumption.}
  {\texttt{TODDLERS\,2.0} provides the flexibility to model diverse stellar populations, cloud structures, and star formation modes within a single framework. It can be used as a standalone tool for individual star-forming regions or as a sub-grid emission model in galaxy-scale simulations.}

{}


\maketitle
\section{Introduction}

The evolution of star-forming regions plays a fundamental role in shaping galactic properties across cosmic time. Stellar feedback from young massive stars, through ionizing radiation, stellar winds, and supernovae, drives the expansion of shells and bubbles into the surrounding interstellar medium (ISM). This feedback regulates subsequent star formation and shapes the observable properties of galaxies \citep{2007ARA&A..45..565M, 2012ARA&A..50..531K, 2014PhR...539...49K}. Modeling these processes self-consistently remains a central challenge in astrophysics.

Three-dimensional radiation-hydrodynamic simulations capture the full complexity of feedback-driven dynamics \citep{2014MNRAS.442..694D, 2018ApJ...859...68K, 2021ApJ...914...90L, 2023MNRAS.521.5160M, 2025ApJ...989...43L} but remain computationally expensive, limiting the range of parameter space that can be explored and making direct coupling to galaxy-scale models impractical for large grids. One-dimensional, spherically symmetric frameworks offer a complementary approach: by solving the equations of motion for an expanding shell driven by stellar feedback, they can efficiently survey broad ranges of cloud masses, densities, metallicities, and stellar population properties. Several groups have developed such frameworks. \citet{2008ApJS..176..438G} use \texttt{MAPPINGS~III} to compute the panchromatic spectral energy distribution (SED) of starburst H\,\textsc{ii} regions as a function of ionization parameter, pressure, and compactness. \texttt{WARPFIELD} \citep{2017MNRAS.470.4453R, 2019MNRAS.483.2547R} and its emission-line extension \texttt{WARPFIELD-EMP} \citep{2020MNRAS.496..339P} couple 1D shell dynamics with \texttt{Cloudy} photoionization modeling to predict observables from feedback-driven H\,\textsc{ii} regions.

The \texttt{TODDLERS} (Time evolution of Observables including Dust Diagnostics and Line Emission from Regions containing young Stars) framework \citep{2023MNRAS.526.3871K, 2024A&A...692A..79K} was developed in this spirit. It operates in two stages: first, a 1D shell evolution module solves the equations of motion for a spherically symmetric shell driven by stellar feedback; second, the resulting time-dependent shell structure is post-processed with the photoionization code \texttt{Cloudy} \citep{2023RMxAA..59..327C} to produce multi-wavelength observables, including full UV-to-mm SEDs, emission-line luminosities, and dust continuum fluxes. The shell evolution module accounts for several feedback mechanisms: mechanical feedback from stellar winds and supernovae, radiation pressure from UV and IR photons, Lyman-$\alpha$ pressure through multiple scatterings, and gravitational forces from the central stellar cluster. The evolution proceeds through several distinct phases, from an initial pressure-driven expansion through fragmentation, followed by momentum-driven expansion and eventual dissolution; if gravity wins the competition against stellar feedback, the shell recollapses and triggers a new generation of star formation. Unlike grids parameterized directly by the local nebular conditions, such as the ionization parameter, gas-phase metallicity, and hydrogen density \citep{2016MNRAS.462.1757G, 2017ApJ...840...44B, 2022ApJ...926...80G, 2023MNRAS.526.3610H, 2025ApJ...986....9L, 2025OJAp....8E.152L}, \texttt{TODDLERS} operates in physical parameter space, with cloud mass, density, stellar mass, metallicity, and star formation history as primary inputs. The local nebular conditions, including the ionization parameter, hydrogen density, dust column, and cloud geometry, are not imposed but emerge self-consistently and time-dependently from the shell evolution. In practice this means that a single \texttt{TODDLERS} model traces a trajectory through the space of local conditions as the region ages, rather than corresponding to a fixed point in it. The predicted observables span from the UV to the mm, covering optical nebular lines, FIR fine-structure lines, and molecular transitions, making them directly comparable to data from \texttt{VLT}/\texttt{MUSE}, \texttt{JWST}/\texttt{NIRSpec} and \texttt{MIRI}, \texttt{CFHT}/\texttt{SITELLE} \citep[e.g.\ the SIGNALS survey;][]{2019MNRAS.489.5530R}, and \texttt{ALMA}. The framework can be used as a standalone tool, to interpret individual star-forming regions, or as a sub-grid emission model in radiative transfer post-processing of galaxy-scale simulations. It has been validated against empirical IR SED templates of star clusters observed with \texttt{HST} and \texttt{JWST} in NGC~628 \citep{2025ApJ...982...50W}.

In its original form \citep[v1;][]{2023MNRAS.526.3871K, 2024A&A...692A..79K}, \texttt{TODDLERS} assumed a fixed stellar population model based on a standard IMF with single-star evolution and instantaneous star formation, a static uniform cloud density, and fixed dust grain properties without accounting for diffuse ionized gas (DIG) emission. Several of these simplifications are challenged by observations and need to be addressed for the framework to be applicable across the range of environments encountered in galaxy simulations.

A fixed IMF and single-star evolution do not capture the full picture: mass-to-light ratios in massive early-type galaxies favor bottom-heavy distributions \citep{2010ARA&A..48..339B}, while top-heavy IMFs have been inferred in high-redshift starbursts and metal-poor systems \citep{2022ApJ...931...57S, 2024ApJ...973L..32V}. A fixed Kroupa or Chabrier IMF cannot capture this diversity. In low-mass clusters ($M_* \lesssim 10^4\,M_\odot$), the IMF is not fully sampled, and the presence or absence of a single very massive star introduces order-of-magnitude variance in the integrated feedback properties \citep{2012ApJ...745..145D}. Binary interactions through mass transfer, common envelope evolution, and stellar mergers can affect both ionizing photon output and spectral hardness on timescales relevant to H\,\textsc{ii} region evolution \citep{2017PASA...34...58E}, motivating the inclusion of binary stellar evolution models. Furthermore, star-forming regions frequently undergo extended periods of star formation rather than instantaneous bursts \citep{2003ARA&A..41...57L}.

The assumption of a static, uniform cloud density is also simplistic: both low-mass pre-stellar cores \citep{2000A&A...361..555B, 2001Natur.409..159A} and massive star-forming clumps \citep{2018A&A...617A.100B} exhibit centrally concentrated density profiles. Pre-supernova feedback, particularly photoionization, has been identified as the primary driver of cloud disruption \citep{2007ARA&A..45..339B, 2022MNRAS.509..272C}, and the surrounding medium responds dynamically to the escaping radiation, an effect not captured by static cloud density models.

The treatment of dust and ionized gas also requires revision. The grain size distribution is not universal: theoretical dust evolution models predict that the small-to-large grain mass ratio varies with local star formation history and ISM conditions \citep{2015MNRAS.447.2937H, 2024A&A...689A..79M, 2026MNRAS.545f2040T, 2026A&A...705A..75M}, affecting not only the continuum attenuation and thermal dust emission but also the nebular line emission through the influence of dust on the ionization structure \citep{1972ApJ...177L..69P, 2000PASJ...52..539I, 2011ApJ...732..100D}. Furthermore, DIG contributes approximately half of the total H$\alpha$ luminosity in nearby star-forming galaxies \citep{2009RvMP...81..969H, 2016MNRAS.461.3111B, 2024MNRAS.532.2016M} and alters emission-line ratios relative to H\,\textsc{ii} region emission alone \citep{2019MNRAS.489.4721V, 2022MNRAS.513.2904T}, yet was not included in the original framework.

Addressing these shortcomings is particularly timely. \texttt{JWST} \texttt{NIRSpec} is delivering optical emission-line spectra across a wide redshift range, from spatially resolved IFU studies of individual star-forming complexes at $z = 4$--$8$ \citep{2024A&A...691A..19V, 2025arXiv251016116F} to large spectroscopic surveys providing integrated emission-line flux catalogs for thousands of galaxies \citep[e.g.][]{2025ApJS..277....4D, 2024A&A...690A...2M, 2025ApJ...983L...4F}. The latter do not resolve individual H\,\textsc{ii} regions and instead capture the superposed emission of many; this is precisely the regime in which \texttt{TODDLERS} is intended to serve as a sub-grid emission model, supplying the building-block spectra that are combined in a galaxy-scale calculation. The systematic offsets seen in BPT diagrams at high redshift \citep{2025ApJ...980..242S} further underscore the need for flexible emission models. To this end, we present \texttt{TODDLERS\,2.0}, which addresses the limitations of the original framework described above, broadening the range of physical conditions that can be modeled.

The remainder of this paper is organized as follows. Sect.~\ref{sect:methodology} describes the new developments in detail. Sect.~\ref{sect:results} presents illustrative test cases that demonstrate the impact of selected features on shell dynamics; a comprehensive parameter study and science exploitation of the full feature set is deferred to a forthcoming paper. Sect.~\ref{sec:summary} provides a summary and outlook for future work.
\section{Framework enhancements}
\label{sect:methodology}

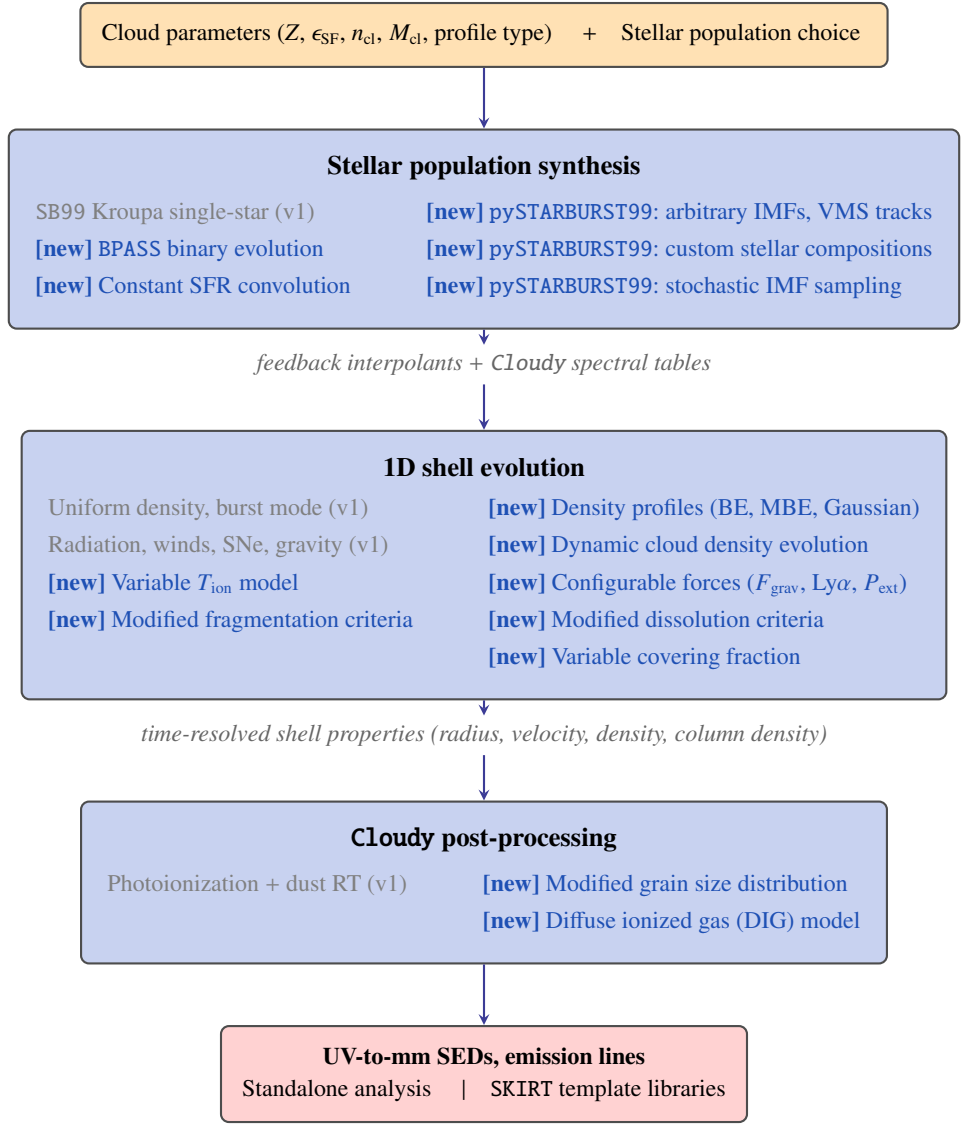
\begin{figure*}
\centering
\definecolor{boxblue}{RGB}{200,210,240}
\definecolor{boxorange}{RGB}{255,225,180}
\definecolor{boxpink}{RGB}{255,210,210}
\definecolor{arrowblue}{RGB}{60,60,150}
\definecolor{newblue}{RGB}{30,80,180}
\begin{tikzpicture}[
    >=stealth,
    stage/.style={
        rectangle, draw=black!70, thick, rounded corners=3pt,
        align=center, inner sep=10pt, fill=boxblue
    },
    iobox/.style={
        rectangle, draw=black!70, thick, rounded corners=3pt,
        align=center, inner sep=8pt, fill=boxorange, font=\small
    },
    outbox/.style={
        rectangle, draw=black!70, thick, rounded corners=3pt,
        align=center, inner sep=8pt, fill=boxpink, font=\small\bfseries
    },
    arw/.style={-{stealth}, thick, arrowblue},
    datalabel/.style={font=\footnotesize\itshape, text=black!60}
]

\node[iobox] (input) {
    Cloud parameters ($Z$, $\epsilon_\mathrm{SF}$, $n_\mathrm{cl}$, $M_\mathrm{cl}$, profile type)
    \quad+\quad Stellar population choice
};

\node[stage, below=0.8cm of input] (stage1) {
    \textbf{Stellar population synthesis}\\[6pt]
    \begin{tabular}{@{}l@{\hspace{1.0cm}}l@{}}
    \textcolor{black!50}{\small \texttt{SB99} Kroupa single-star (v1)} &
    \textcolor{newblue}{\small \textbf{[new]} \texttt{pySTARBURST99}: arbitrary IMFs, VMS tracks} \\[3pt]
    \textcolor{newblue}{\small \textbf{[new]} \texttt{BPASS} binary evolution} &
    \textcolor{newblue}{\small \textbf{[new]} \texttt{pySTARBURST99}: custom stellar compositions} \\[3pt]
    \textcolor{newblue}{\small \textbf{[new]} Constant SFR convolution} &
    \textcolor{newblue}{\small \textbf{[new]} \texttt{pySTARBURST99}: stochastic IMF sampling} \\
    \end{tabular}
};

\node[datalabel, below=0.2cm of stage1] (df1) {feedback interpolants + \texttt{Cloudy} spectral tables};

\node[stage, below=0.6cm of df1] (stage2) {
    \textbf{1D shell evolution}\\[6pt]
    \begin{tabular}{@{}l@{\hspace{1.0cm}}l@{}}
    \textcolor{black!50}{\small Uniform density, burst mode (v1)} &
    \textcolor{newblue}{\small \textbf{[new]} Density profiles (BE, MBE, Gaussian)} \\[3pt]
    \textcolor{black!50}{\small Radiation, winds, SNe, gravity (v1)} &
    \textcolor{newblue}{\small \textbf{[new]} Dynamic cloud density evolution} \\[3pt]
    \textcolor{newblue}{\small \textbf{[new]} Variable $T_\mathrm{ion}$ model} &
    \textcolor{newblue}{\small \textbf{[new]} Configurable forces ($F_\mathrm{grav}$, Ly$\alpha$, $P_\mathrm{ext}$)} \\[3pt]
    \textcolor{newblue}{\small \textbf{[new]} Modified fragmentation criteria} &
    \textcolor{newblue}{\small \textbf{[new]} Modified dissolution criteria} \\[3pt]
    &
    \textcolor{newblue}{\small \textbf{[new]} Variable covering fraction} \\
    \end{tabular}
};

\node[datalabel, below=0.2cm of stage2] (df2) {time-resolved shell properties (radius, velocity, density, column density)};

\node[stage, below=0.6cm of df2] (stage3) {
    \textbf{\texttt{Cloudy} post-processing}\\[6pt]
    \begin{tabular}{@{}l@{\hspace{1.0cm}}l@{}}
    \textcolor{black!50}{\small Photoionization + dust RT (v1)} &
    \textcolor{newblue}{\small \textbf{[new]} Modified grain size distribution} \\[3pt]
    &
    \textcolor{newblue}{\small \textbf{[new]} Diffuse ionized gas (DIG) model} \\
    \end{tabular}
};

\node[outbox, below=0.8cm of stage3] (output) {
    UV-to-mm SEDs, emission lines\\[2pt]
    \normalfont Standalone analysis \quad$\vert$\quad \texttt{SKIRT} template libraries
};

\draw[arw] (input) -- (stage1);
\draw[arw] (stage1) -- (df1);
\draw[arw] (df1) -- (stage2);
\draw[arw] (stage2) -- (df2);
\draw[arw] (df2) -- (stage3);
\draw[arw] (stage3) -- (output);

\end{tikzpicture}
\caption{Schematic overview of the \texttt{TODDLERS\,2.0} framework. Features introduced in this work are marked in blue; existing capabilities from the original framework are shown in gray. Italic labels indicate the data passed between processing steps.}
\label{fig:flowchart}
\end{figure*}

This section describes the enhancements to the \texttt{TODDLERS} framework. Figure~\ref{fig:flowchart} and Table~\ref{tab:new_features} provide an overview. We have verified that the updated code reproduces the v1 results when run with identical input parameters, confirming full backwards compatibility.

\begin{table*}
\caption{New features and enhancements in \texttt{TODDLERS\,2.0} relative to the original framework \citep{2023MNRAS.526.3871K}.}
\label{tab:new_features}
\centering
\begin{tabular}{llll}
\hline\hline
Feature & Sect. & Key parameter(s) & Default \\
\hline
Constant SFR mode           & \ref{subsect:constant_sfr}         & Formation timescale $\tau_\mathrm{SF}$              & Off (burst) \\
\texttt{BPASS} binary evolution & \ref{subsubsect:binary_evolution}  & $M_\mathrm{upper}$ (100 or 300\,$M_\odot$)          & Optional \\
\texttt{pySTARBURST99}      & \ref{subsect:pysb99}               & IMF shape, $M_\mathrm{upper}$                       & Optional \\
Stochastic IMF sampling     & \ref{subsect:stochastic_sampling}  & —                                                   & Optional \\
Arbitrary density profiles  & \ref{subsect:density_profiles}     & Power-law slope $\alpha$ (MBE)                      & Uniform \\
Dynamic cloud density       & \ref{subsect:dynamic_density}      & $\eta_\mathrm{KE} = 7.5\times10^{-4}$              & Off \\
Modified fragmentation      & \ref{subsect:fragmentation}        & $\alpha_\mathrm{sc}$, $t_\mathrm{min}$, $\rho_\mathrm{thresh}$ & Density contrast + time req. \\
Modified dissolution        & \ref{subsect:dissolution}          & Swept-mass fraction $k$                             & $k = 1$ \\
Variable covering fraction  & \ref{subsect:covering_fraction}    & $f_\mathrm{cover,min}$                              & Off ($f_\mathrm{cover,min} = 1$) \\
\multirow{3}{*}{Configurable force terms} & \multirow{3}{*}{\ref{app:force_terms}} & Gravity ($F_\mathrm{grav}$)        & On \\
                            &                                    & Lyman-$\alpha$ ($F_{\mathrm{Ly}\alpha}$)            & On \\
                            &                                    & External pressure ($P_\mathrm{ext}$)                & On \\
Variable $T_\mathrm{ion}$   & \ref{app:variable_tion}        & $(Z,\,n_\mathrm{H},\,\phi)$ grid                   & Off ($10^4$\,K fixed) \\
Modified grain distribution & \ref{subsubsect:grain_size}        & Small/large grain mass ratio                        & 0.1 \\
Diffuse ionized gas         & \ref{subsubsect:dig}               & $\log U_\mathrm{bg}$, $n_\mathrm{DIG}$             & Off \\
\hline
\end{tabular}
\end{table*}

\subsection{Constant star formation rate mode}
\label{subsect:constant_sfr}

Star formation in molecular clouds is observed to proceed over extended timescales rather than in a single instantaneous event \citep{2003ARA&A..41...57L}. We implement a constant star formation rate (SFR) mode in addition to the instantaneous burst mode from the original framework, distributing star formation over a specified timescale $\tau_{\mathrm{SF}}$. The stellar feedback quantities in constant SFR mode are calculated through temporal convolution of the burst mode results. For any feedback quantity $Q(t)$ (such as mechanical luminosity or ionizing photon rate), the constant SFR result is computed as

\begin{equation}
    Q_{\mathrm{CSF}}(t) = \int_0^{\min(t,\tau_{\mathrm{SF}})} \tilde{Q}_{\mathrm{burst}}(t-\tau) \frac{M_*}{\tau_{\mathrm{SF}}} d\tau,
\end{equation}
where $\tilde{Q}_{\mathrm{burst}}(t)$ is the feedback quantity per unit stellar mass for an instantaneous burst (normalized to $1\,M_\odot$), and $M_*/\tau_{\mathrm{SF}}$ is the star formation rate. The integrand therefore gives the contribution from a mass element $\mathrm{d}m = (M_*/\tau_{\mathrm{SF}})\,\mathrm{d}\tau$ formed at time $\tau$. This formulation ensures a smooth evolution of feedback properties over time, better matching the observed characteristics of star-forming regions with extended periods of star formation.

\begin{figure}
    \centering
    \includegraphics[width=0.925\linewidth]{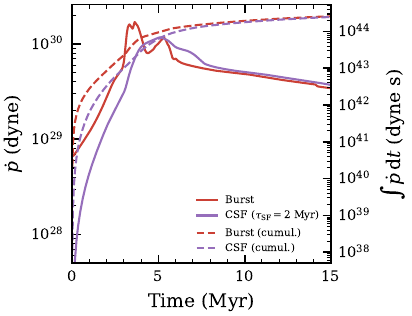}
    \caption{Comparison of ram force evolution between instantaneous burst (red) and constant SFR (purple) modes for identical total stellar mass. The ram force $\dot{p}$ captures momentum injection from mass loss (stellar winds and supernovae). Solid lines show the instantaneous $\dot{p}$ (left axis), while dashed lines show its time integral (right axis). The burst mode produces an immediate peak followed by a decline punctuated by supernova contributions, whereas the constant SFR mode exhibits a gradual rise to a plateau as successive stellar generations contribute. Despite the different temporal profiles, both modes converge to similar cumulative momentum input by $\sim$15~Myr.}
    \label{fig:constant_sfr_vs_burst}
\end{figure}

In practice, we discretize this integral with a time step of 0.025 Myr, which provides sufficient temporal resolution while maintaining computational efficiency. The implementation scales linearly with the formation timescale and preserves total energy and momentum input for a given stellar mass. Figure \ref{fig:constant_sfr_vs_burst} compares the evolution of ram force for burst and constant SFR modes with identical total stellar mass, illustrating the more gradual rise and extended plateau characteristic of continuous star formation. The impact on shell evolution is demonstrated in Sect.~\ref{subsect:profile_sf_comparison}.

\subsection{Stellar population synthesis}
\label{subsect:stellar_populations}

Evidence for IMF variations across environments and the importance of binary evolution for the ionizing output of stellar populations motivate a more flexible treatment of stellar populations in \texttt{TODDLERS}. We now support two independent frameworks: binary stellar evolution through \texttt{BPASS} and a flexible single/multi-star approach through \texttt{pySTARBURST99}. Both frameworks feed into the same convolution infrastructure (Sect.~\ref{subsect:constant_sfr}) and produce the same set of output quantities used in the dynamical evolution: ionizing photon rates, bolometric luminosities, stellar wind mechanical power, and spectral energy distributions for \texttt{Cloudy} post-processing. The two pathways are complementary, and the choice between them depends on the physical scenario and the desired flexibility in stellar population properties.

\subsubsection{Binary stellar evolution}
\label{subsubsect:binary_evolution}

We have incorporated binary stellar evolution models through the \texttt{BPASS} synthesis code \citep[currently v2.2.1;][]{2017PASA...34...58E, 2018MNRAS.479...75S}, which accounts for the effects of binary interactions on stellar evolution and feedback properties. The modular design of the interface ensures that newer \texttt{BPASS} releases can be adopted with minimal effort. Binary stars can significantly alter the ionizing photon output and spectral characteristics. At ages beyond ${\sim}10$~Myr, binary evolution pathways produce hot Wolf-Rayet and helium stars that sustain strong ionizing flux long after single-star models have faded; stars stripped of their envelopes through binary interaction dominate the ionizing emission at these later ages with harder spectra than their massive-star progenitors, affecting the nebular ionization structure \citep{2019A&A...629A.134G}. These differences propagate into emission-line diagnostics: \citet{2018MNRAS.477..904X} show that binary populations maintain high \OIII/H$\beta$ ratios and shift BPT diagram predictions relative to single-star models when compared against observed H\,\textsc{ii} regions. At sub-solar metallicities the ionizing flux boost from binary pathways can reach ${\sim}60$\%, with implications for maintaining cosmic reionization \citep{2016MNRAS.456..485S}.

The mechanical feedback driving the shell dynamics is obtained directly from the \texttt{BPASS} population outputs, which tabulate the wind and supernova energy-injection rates and the corresponding mass-loss rates as functions of population age. We take the mechanical luminosity to be the sum of the wind and supernova energy rates, and obtain the ram (momentum) force from the energy and mass-loss rates via $\dot{p} = \sqrt{2\,\dot{E}\,\dot{M}}$, evaluated separately for the wind and supernova channels and summed.

The implementation includes \citet{2003PASP..115..763C} initial mass functions (IMFs) with upper limits of 100~$M_{\odot}$ and 300~$M_{\odot}$, and 13 metallicities spanning $Z = 10^{-5}$ to $Z = 0.04$. Because the public \texttt{BPASS} databases release IMF-integrated, population-level feedback quantities for this fixed set of IMFs, custom IMF specification and stochastic stellar population sampling (Sect.~\ref{subsect:stochastic_sampling}) are not available for \texttt{BPASS} models; those capabilities make use of \texttt{pySTARBURST99}, which exposes individual stellar tracks and arbitrary stellar compositions.

\subsubsection{pySTARBURST99 stellar population synthesis}
\label{subsect:pysb99}

We have integrated \texttt{pySTARBURST99} \citep{Hawcroft2025}, a Python-based successor to the widely-used \texttt{STARBURST99} stellar population synthesis code \citep{1999ApJS..123....3L}, into the \texttt{TODDLERS} framework. \texttt{pySTARBURST99} provides access to individual stellar evolutionary tracks, enabling the flexible population modeling described below.
The \texttt{pySTARBURST99} integration provides the following capabilities:
\begin{enumerate}
    \item The code incorporates both rotating and non-rotating Geneva evolutionary tracks with stellar masses extending beyond the previous $120\,M_\odot$ limit. This extended mass range increases both ionizing flux and wind momentum predictions during the first few Myr of evolution \citep{Hawcroft2025}.

    \item Users can define arbitrary power-law IMFs with multiple segments by specifying the exponents and mass limits. For example, a Kroupa IMF can be specified with exponents $\alpha_1 = 1.3$ for $0.1 < M/M_{\odot} < 0.5$ and $\alpha_2 = 2.3$ for $0.5 < M/M_{\odot} < M_{\mathrm{upper}}$. The interface is similarly well-suited to accommodate non-standard IMF forms motivated by observations, such as the multi-segment forms proposed for massive galaxies and high-redshift stellar populations \citep{2024ApJ...973L..32V, 2022ApJ...931...57S}.

    \item Beyond IMF-sampled populations, users can specify exact stellar compositions by providing the number of stars at each mass. This capability is particularly useful for modeling individual stellar associations or for investigating the effects of specific massive stars on feedback properties. Custom populations can be scaled to any total stellar mass, allowing for flexible exploration of different cluster masses.
    This framework can also be used to generate stochastic stellar populations as described in Sect.~\ref{subsect:stochastic_sampling}.

\end{enumerate}

The available metallicity grid spans six values: $Z = 0$ (primordial), $0.0004$, $0.002$, $0.006$, $0.014$, and $0.020$. The maximum stellar mass in the evolutionary tracks depends on metallicity: up to $500\,M_\odot$ (non-rotating) at $Z = 0.014$, up to $300\,M_\odot$ at $Z = 0.006$, $0.020$, and $0$ (primordial), and $120\,M_\odot$ at $Z = 0.002$ and $0.0004$. The \texttt{pySTARBURST99} feedback quantities can be pre-computed on a time grid of choice (only limited by the intrinsic time-resolution of the \texttt{pySTARBURST99} tabulated data) and stored as interpolation tables required by the \texttt{TODDLERS} framework. We also generate stellar continuum tables in the format required by \texttt{Cloudy}, enabling self-consistent photoionization modeling with the same stellar populations used in the dynamical evolution. Illustrative applications of the IMF flexibility and custom population capability are presented in Sects.~\ref{subsect:imf_comparison} and~\ref{subsect:custom_population}.

\subsection{Stochastic stellar population sampling}
\label{subsect:stochastic_sampling}

For low-mass stellar populations where the fully-sampled IMF assumption breaks down, we have implemented a stochastic sampling mode that explicitly samples individual stars from the IMF. This approach is particularly important for modeling regions with total stellar masses below $\sim 10^4$~$M_{\odot}$, where statistical fluctuations in the massive star population can significantly affect the integrated feedback properties.

The implementation draws individual stellar masses from any user-specified probability density function $\xi(m)$ until the cumulative mass exceeds a prescribed target $M_*$, then truncates the sample at the star that crosses the threshold (the ``stop-after'' method in the terminology of \citealt{2010A&A...512A..79H}). Each continuous mass is then rounded to the nearest entry in the discrete grid of evolutionary tracks available for the chosen metallicity, so that all feedback quantities are read directly from the pre-computed single-star database without interpolation in mass. An upper stellar mass limit $M_\mathrm{upper}$ truncates the track grid before sampling, ensuring that the full mass budget is distributed among stars with $m \leq M_\mathrm{upper}$. We tested this procedure with 500 independent realizations at $M_* = 10^3\,M_\odot$ and find a median absolute deviation of the total sampled mass from the target below 1 per cent, though individual realizations can deviate by up to several per cent when the last star drawn is massive. We provide several common IMFs as defaults (Kroupa, Salpeter, Chabrier), but custom distributions including broken power laws with arbitrary slopes and break points can also be specified.

For each sampled stellar population, stars are assigned formation times $\tau_i$ drawn from the star formation history. For an instantaneous burst, all stars form at $\tau_i = 0$, while for constant star formation over a timescale $\tau_{\mathrm{SF}}$, formation times are drawn uniformly from $[0, \tau_{\mathrm{SF}}]$, independently of stellar mass. We thus do not impose a mass-dependent formation sequence; all masses are equally likely to form at any time within $\tau_{\mathrm{SF}}$. At any simulation time $t$, a star formed at time $\tau_i$ has stellar age $(t - \tau_i)$ and contributes feedback only for $t \geq \tau_i$. The feedback quantity $Q$ for the population at simulation time $t$ is computed by summing over all stars that have formed
\begin{equation}
    Q_{\mathrm{total}}(t) = \sum_{i: t \geq \tau_i} Q_{\mathrm{single}}(m_i, t - \tau_i),
\end{equation}
where $Q_{\mathrm{single}}(m, t_{\mathrm{age}})$ is the feedback from an individual star of mass $m$ at stellar age $t_{\mathrm{age}}$, obtained from a pre-computed database of evolutionary tracks generated with \texttt{pySTARBURST99}, making use of the ability to retrieve exact stellar compositions as described in Sect.~\ref{subsect:pysb99}. Binary populations are not supported in stochastic mode (see Sect.~\ref{subsubsect:binary_evolution}).
For each stochastic realization, we generate both feedback tables and SEDs, just as for fully-sampled populations, to enable self-consistent evolution and photoionization modeling. The impact of stochastic sampling on shell dynamics is demonstrated in Sect.~\ref{subsect:stochastic_results}.

\subsection{Mass assignment in the stochastic mode}
\label{subsect:mass_assignment}

Each stochastically sampled star draws its feedback from the single-star database, which is tabulated on a discrete grid of evolutionary tracks (Sect.~\ref{subsect:pysb99}). Interpolating this feedback in mass at fixed clock age is unphysical: stars of different mass reach a given evolutionary stage (for example the Wolf-Rayet phase) at different ages, so fixed-age interpolation averages stars in unlike phases and smears the sharp transitions in their feedback. We therefore snap each sampled star to the nearest grid mass by default, which assigns it a real, self-consistent evolutionary history.

\begin{figure*}
\centering
\includegraphics[width=0.85\linewidth]{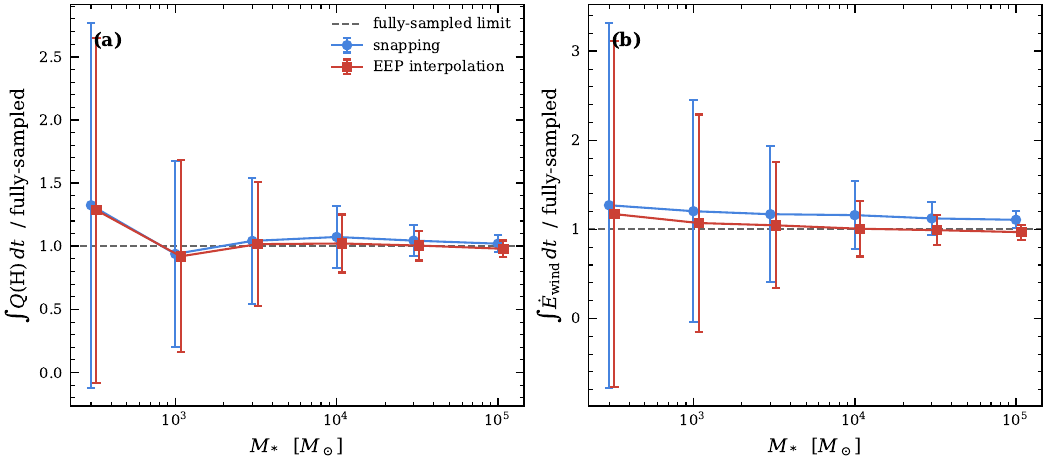}
\caption{Mass assignment in the stochastic mode: time-integrated ionizing-photon (panel a) and wind-power (panel b) budgets of bursts sampled from a Kroupa IMF ($0.1$--$100\,M_\odot$), normalized to the fully-sampled limit (dashed), as a function of stellar mass $M_*$, for nearest-grid snapping (blue) and EEP interpolation (red). Error bars show the realization-to-realization scatter (30 realizations per point, MW metallicity). In the under-sampled regime the scatter dominates the difference between the two methods; at full sampling, snapping mildly overestimates the wind budget, which EEP interpolation removes.}
\label{fig:snap_eep}
\end{figure*}

To quantify the cost of this choice, we have also implemented the physically motivated alternative, interpolation along Equivalent Evolutionary Points (EEPs), which aligns the evolutionary phase before interpolating (Appendix~\ref{app:eep}). Figure~\ref{fig:snap_eep} compares the two, showing the time-integrated ionizing-photon and wind-power budgets of bursts sampled from a Kroupa IMF (using the stop-after method of Sect.~\ref{subsect:stochastic_sampling}), normalized to the fully-sampled limit, as a function of stellar mass $M_*$. A small systematic offset between the two methods is present at all stellar masses: the convex feedback-mass relation makes nearest-grid snapping mildly overestimate the budget, by about 10 to 15 per cent for the wind power and a few per cent for the ionizing output. In the under-sampled regime ($M_* \lesssim 10^4\,M_\odot$) for which stochastic sampling is intended, however, this offset is buried under the realization-to-realization scatter, which is of order 100 per cent, so the mass-assignment choice is immaterial there. The offset becomes relevant only once the IMF is well sampled and the scatter collapses; EEP interpolation removes it and recovers the fully-sampled limit. The effect is expected to scale with the coarseness of the high-mass track grid.

\subsection{Cloud density profiles}
\label{subsect:density_profiles}

We have extended \texttt{TODDLERS} to support physically motivated density distributions beyond the uniform density assumption of the original framework. Observations across a wide range of cloud masses show that molecular clouds are not uniform: extinction mapping of nearby cores reveals Bonnor-Ebert-like radial profiles \citep{2000A&A...361..555B, 2001Natur.409..159A, 2007ARA&A..45..339B}, and similar centrally peaked structures are found in massive clumps hosting high-mass star formation \citep{2003ApJ...592..188B, 2018A&A...617A.100B}. Our implementation enables users to select from several profile types while maintaining consistent treatment of shell dynamics.

Particularly useful is the modified Bonnor-Ebert (MBE) profile, which implements a piecewise density structure that maintains a constant central density $\rho_0$ up to radius $R_0$, followed by a power-law decline with exponent $\alpha$ in the outer regions

\begin{equation}
    \rho(r) = \begin{cases}
        \rho_0 & r \leq R_0 \\
        \rho_0(r/R_0)^{-\alpha} & R_0 < r \leq R_{\mathrm{cl}} \\
        \rho_{\mathrm{amb}} & r > R_{\mathrm{cl}}.
    \end{cases}
\label{eqn:MBE_equation}
\end{equation}

We determine the transition radius $R_0$ numerically to ensure mass conservation for the given cloud mass $M_{\mathrm{cl}}$ and average cloud density $n_{\mathrm{cl}}$. The parameter $\alpha$ controls the steepness of the density decline and can be adjusted to match specific cloud observations, with typical values between 1.5 and 2.5. For all profiles, we calculate the gravitational binding energy numerically using integration of the gravitational potential energy density.

Beyond the modified Bonnor-Ebert profile, our framework also supports:
\begin{enumerate}
    \item Standard Bonnor-Ebert spheres, solved using the Lane-Emden equation for isothermal, self-gravitating spheres in pressure equilibrium
    \item Gaussian profiles, which follow an exponential decline with radius squared
\end{enumerate}

\begin{figure}
    \centering
    \includegraphics[width=0.925\linewidth]{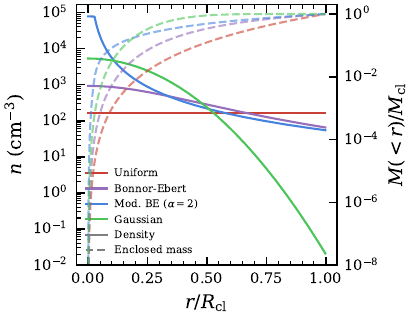}
    \caption{Radial number density (solid lines, left axis) and enclosed mass fraction (dashed lines, right axis) for the four cloud density profiles implemented in \texttt{TODDLERS\,2.0}: uniform (red), Bonnor-Ebert (purple), modified Bonnor-Ebert (blue), and Gaussian (green). All profiles share the same total cloud mass ($\log\,M_\mathrm{cl}/M_\odot = 6.0$) and average density ($n_\mathrm{cl} = 160\,\mathrm{cm}^{-3}$), matching the parameters used in Sect.~\ref{subsect:profile_sf_comparison}.}
    \label{fig:density_profiles}
\end{figure}

Figure~\ref{fig:density_profiles} compares the radial density and enclosed mass fraction for the implemented profiles. The implementation of these varied profiles allows investigation of how cloud structure affects feedback-driven shell evolution. Our numerical framework automatically adapts initial conditions based on the selected profile type, calculating the appropriate starting radius and velocity using either analytical solutions (for uniform density) or numerical root-finding procedures (for non-uniform profiles). The density profile implementation is modular: each profile provides a common interface for calculating density, mass, and binding energy, so adding a new profile requires only implementing these methods. The effect of cloud structure on shell evolution is explored in Sect.~\ref{subsect:profile_sf_comparison}.

\subsection{Dynamic cloud density evolution}
\label{subsect:dynamic_density}

Pre-supernova feedback, particularly photoionization, has been identified as the primary mechanism disrupting molecular clouds \citep{2022MNRAS.509..272C}, and the surrounding medium is expected to respond dynamically to the injected energy. Whereas Sect.~\ref{subsect:density_profiles} concerns the initial spatial structure of the cloud, this section addresses whether the mean density of the unswept cloud changes over the course of the simulation in response to stellar feedback. We implement dynamic evolution of the cloud density in response to stellar feedback, an important capability when significant energy is injected into the unswept portion of the cloud. The broader question of how stellar winds and ionizing radiation interact self-consistently in a feedback bubble has been explored analytically and numerically by \citet{2025ApJ...989...42L, 2025ApJ...989...43L}, who model the coevolution of the wind-blown bubble and photoionized region interior to the expanding shell, including the effect of the photoionized region on cooling at the wind bubble interface. The treatment described here is complementary but distinct: rather than focusing on the internal bubble structure, we model how radiation escaping through the outer shell heats and expands the unswept cloud exterior to it, reducing back-pressure on the shell from the outside. We emphasize that this is an additional, external channel: the shell itself is always driven by the full equation of motion (Appendix~\ref{app:force_terms}), which combines the hot-bubble thermal pressure, stellar-wind and supernova ram pressure, UV, IR, and Lyman-$\alpha$ radiation pressure, gravity, and external confining pressure, with the dynamic cloud density entering only through the external-pressure term. A systematic comparison between these two approaches, and their combined effect on shell dynamics, is a natural direction for future work. Currently, this functionality is implemented for uniform density profiles only. We model the cloud as a uniform density sphere with a central cavity, where the cavity radius corresponds to the position of the expanding shell. As the shell expands, it defines the boundary between the swept-up material (contained in the shell) and the remaining unswept cloud.

The dynamic density evolution is based on energy conservation principles. When energy is injected into the unswept cloud, this modifies its binding energy, resulting in expansion. For an energy injection $E_{\mathrm{inj}}$, we determine the new cloud configuration by solving
\begin{equation}
U_{\mathrm{final}} = U_{\mathrm{initial}} + E_{\mathrm{inj}}.
\end{equation}
For a uniform density cloud with a central cavity, the binding energy is given by
\begin{equation}
U = -\frac{3GM_{\mathrm{gas}}^2}{5R_{\mathrm{cl}}}\left(1-\left(\frac{R_{\mathrm{cav}}}{R_{\mathrm{cl}}}\right)^5\right),
\end{equation}
where $M_{\mathrm{gas}}$ is the mass of the unswept cloud material, $R_{\mathrm{cl}}$ is the cloud radius, and $R_{\mathrm{cav}}$ is the cavity radius (equal to the shell radius). When energy is injected, the cloud radius increases while preserving the total gas mass and cavity radius, resulting in a decreased average density.
The energy injection rate is calculated from the ionizing radiation that escapes the shell and is absorbed by the unswept cloud,
\begin{equation}
\dot{E}_{\mathrm{inj}} = (1 - f_{\mathrm{esc,cloud}}) \, f_{\mathrm{esc,shell}} \, \eta_{\mathrm{KE}} \, Q(\mathrm{H}) \, E_{\mathrm{LyC}},
\end{equation}
where $f_{\mathrm{esc,shell}}$ is the escape fraction of ionizing photons through the shell, $f_{\mathrm{esc,cloud}}$ is the escape fraction from the unswept cloud, $\eta_{\mathrm{KE}}$ is the kinetic energy efficiency parameter (set to $7.5 \times 10^{-4}$ based on \citealt{2003ApJ...594..888F}), $Q(\mathrm{H})$ is the rate of photons with energy above 13.6\,eV (the hydrogen ionization threshold), and $E_{\mathrm{LyC}}$ is the average energy of ionizing photons. Both escape fractions are computed during the 1D shell evolution by integrating the ionizing photon survival probability through the shell and cloud respectively, accounting for losses to both hydrogen recombination and dust absorption.
A potential impact of this cloud expansion is the reduction in back pressure on the expanding shell. As the unswept cloud density $n$ decreases, the thermal pressure ($P \propto n \, T$) opposing shell expansion diminishes, potentially allowing for more efficient shell acceleration. This effect may be pronounced in ionized regions where the temperature $T_{\mathrm{ion}} \approx 10^4$ K creates substantial pressure. By modeling this density-dependent back pressure, we explore another possible feedback channel.
This mechanism could be particularly important at low metallicities ($Z \lesssim 0.008 Z_{\odot}$) due to the lower metal content and reduced dust opacity while simultaneously weakening stellar winds. This makes radiation the dominant feedback channel with enhanced penetration into the cloud. Our implementation captures this metallicity dependence through the calculation of escape fractions and energy conversion efficiencies. This interconnected evolution between the shell and ambient medium provides additional insights when modeling feedback processes in metal-poor environments relevant to high-redshift galaxies and local dwarf galaxies. A demonstration of this effect is presented in Sect.~\ref{subsect:dynamic_density_results}.

\subsection{Fragmentation criteria}
\label{subsect:fragmentation}
As the expanding shell sweeps up ambient material, it can fragment through Rayleigh-Taylor instability, when the shell is accelerated into a lower-density environment, or through gravitational instability, when the shell's self-gravity exceeds its thermal and kinetic support. Fragmentation triggers the transition from pressure-driven to momentum-driven expansion \citep[see][for a detailed discussion]{2023MNRAS.526.3871K}. In the original framework, the acceleration-based fragmentation was determined solely using the pressure acceleration criterion (see $\beta$ condition below). We have implemented a more flexible approach that allows for multiple criteria.
The acceleration-based fragmentation now incorporates the following options:
\begin{enumerate}
\item Beta condition: This criterion requires that $\beta \leq 0$, where $\beta = -(t/P)\,(dP/dt)$. Since $\beta$ is the negative of the relative pressure derivative, a negative $\beta$ value implies that the internal pressure is increasing.
\item Density contrast: This criterion requires the shell density to exceed the external density by at least a factor of $\rho_{\mathrm{shell}}/\rho_{\mathrm{external}} \geq \rho_{\mathrm{thresh}}$, where $\rho_{\mathrm{thresh}}$ is a configurable parameter that defaults to 1.1.
\end{enumerate}
These criteria can be enabled independently or in combination, allowing for greater flexibility in modeling different physical scenarios. The gravitational fragmentation condition, which triggers when the shell's gravitational binding energy exceeds its thermal and kinetic energy \citep[equation~8 in][]{2023MNRAS.526.3871K}, remains unchanged from the original framework.
We also introduce a time duration requirement for the fragmentation model. For any acceleration-based criterion to trigger fragmentation, the condition must persist for a minimum time period that scales with the shell's sound crossing time
\begin{equation}
t_{\mathrm{criterion}} > \max(t_{\mathrm{min}}, \alpha_{\mathrm{sc}} \cdot t_{\mathrm{sound}}),
\end{equation}
where $t_{\mathrm{sound}} = d_{\mathrm{shell}}/c_s$ is the sound crossing time across the shell thickness $d_{\mathrm{shell}}$, $\alpha_{\mathrm{sc}}$ is a configurable parameter (typically set to 1.0), and $t_{\mathrm{min}}$ is a minimum threshold (typically 0.1 Myr) to avoid premature fragmentation due to numerical fluctuations. This ensures that only physically meaningful instabilities trigger fragmentation, as the sound crossing time represents the characteristic timescale for pressure waves to equilibrate across the shell.
For the sound speed, we calculate a thickness-weighted average that accounts for both ionized and neutral components of the shell
\begin{equation}
c_s = w_{\mathrm{ion}} \cdot c_{s,\mathrm{ion}} + w_{\mathrm{neutral}} \cdot c_{s,\mathrm{neutral}},
\end{equation}
where $w_{\mathrm{ion}}$ and $w_{\mathrm{neutral}}$ are the fractions of the shell thickness in ionized and neutral states, while $c_{s,\mathrm{ion}}$ and $c_{s,\mathrm{neutral}}$ are the corresponding sound speeds.

\subsection{Dissolution criteria} 
\label{subsect:dissolution}
In the original \texttt{TODDLERS} framework, we implemented multiple dissolution criteria where the shell was considered dissolved when: (1) its maximum density fell below 0.5 cm$^{-3}$ for a period exceeding 1 Myr, (2) the shell velocity remained below a minimum value (shell stalling) for an extended duration, or (3) the shell expanded beyond a maximum radius (1000 pc). These criteria were applied only after the entire cloud had been swept up.

In the \texttt{TODDLERS\,2.0} framework, we have maintained these core criteria while adding new capabilities specifically related to dynamic cloud density evolution. The dissolution phase continues to represent the final stage of shell evolution, where the shell becomes too diffuse to be treated as a coherent structure. For our newly implemented dynamic cloud density models, the system is considered to be dissolved when the unswept cloud material has expanded to the ambient ISM density (0.1 cm$^{-3}$), provided either sufficient mass has been swept up ($M_\mathrm{sh} \geq k \, M_\mathrm{cl}$, with $k = 1$ by default) or the shell has stalled for an extended period.

This criterion accounts for scenarios where the unswept portion of the cloud has dynamically expanded to ISM densities. In such cases, the physical distinction between the cloud and the ambient medium becomes increasingly blurred, justifying the dissolution transition. As in Paper I, we maintain the time duration requirement (1 Myr) for all criteria to prevent transient fluctuations from prematurely triggering dissolution.

\subsection{Covering fraction}
\label{subsect:covering_fraction}

We have implemented a more sophisticated treatment of the shell covering fraction. In the original framework, the shell was assumed to fully cover the central source throughout the evolution. In this version, the covering fraction remains unity while the shell is still sweeping up cloud material ($M_\mathrm{sh} < M_\mathrm{cl}$). Once the cloud is fully swept ($M_\mathrm{sh} \approx M_\mathrm{cl}$), $f_{\mathrm{cover}}$ transitions smoothly from 1 to a user-specified floor value $f_{\mathrm{cover,min}}$ over a timescale of 0.25~Myr, using a hyperbolic tangent profile in time. If the shell subsequently recollapses and falls back within half the original cloud radius ($R < 0.5\,R_\mathrm{cl}$), $f_{\mathrm{cover}}$ transitions smoothly back to unity via a hyperbolic tangent profile in $R/R_{\mathrm{cl}}$.

This implementation affects the radiative and mechanical forces on the shell, providing a more realistic representation of the holes in the shell that cannot be directly modeled in a 1D framework. We note that the covering fraction and the ionizing photon escape fraction are distinct quantities in our framework: a non-zero escape fraction can occur even with $f_\mathrm{cover} = 1$ when the shell becomes optically thin (Sect.~\ref{subsect:dynamic_density}), while $f_\mathrm{cover} < 1$ represents geometric holes through which radiation escapes without interacting with the shell at all. The parameter $f_{\mathrm{cover,min}}$ is a free parameter whose value depends on the geometry of the specific region being modeled. Observational constraints on the porosity of individual H\,\textsc{ii} region shells remain limited, though the observed escape fraction of ionizing photons from resolved star-forming regions provides an indirect proxy for the combined effect of geometric and transmitted escape of ionizing radiation. The covering fraction also controls the fraction of direct stellar radiation that reaches the diffuse ionized gas component (Sect.~\ref{subsubsect:dig}).

Two further refinements to the shell dynamics are worth noting before turning to the \texttt{Cloudy} post-processing. The momentum equation includes several force terms (thermal pressure, radiation pressure, stellar wind ram pressure, gravity, external pressure, and momentum loading from swept-up mass) that can be individually activated or deactivated, allowing users to isolate the contribution of each physical process. The detailed formalism, including the Lyman-$\alpha$ resonant scattering model, is described in Appendix~\ref{app:force_terms}.

In the original framework, the temperature of the ionized gas at the shell's inner face was fixed at $T_{\mathrm{ion}} = 10^4$~K. We now provide an option to compute this temperature self-consistently from a pre-calculated grid of \texttt{Cloudy} photoionization models, parameterized by metallicity, hydrogen density, and ionizing photon flux (Appendix~\ref{app:variable_tion}).

\subsection{Modifications to \texttt{Cloudy} post-processing}
\label{subsect:cloudy_mods}

As illustrated in Fig.~\ref{fig:flowchart}, the \texttt{TODDLERS} framework operates in two stages: the 1D shell evolution module tracks the dynamical response of the cloud to stellar feedback, while the resulting time-dependent shell properties are post-processed with \texttt{Cloudy} to produce multi-wavelength observables. The preceding sections focused on updates to the shell evolution module; here we describe modifications to the \texttt{Cloudy} post-processing stage.

\begin{table*}[t!]
\caption{Physical parameters for the simulation sets presented in Sect.~\ref{sect:results}.}
\label{tab:sim_params}
\centering
\begin{tabular}{lcccccc}
\hline\hline
Section (Figure) & $Z$ & $n_\mathrm{cl}$\,[cm$^{-3}$] & $\log(M_\mathrm{cl,init}/M_\odot)$ & $\epsilon_\mathrm{SF}$ & Mode & Profile \\
\hline
\multirow{4}{*}{\ref{subsect:profile_sf_comparison} (Fig.~\ref{fig:profile_sf_comparison})}
    & \multirow{4}{*}{0.014} & \multirow{4}{*}{160} & \multirow{4}{*}{6.0} & \multirow{4}{*}{0.05}
    & Burst & Uniform \\
 & & & & & Burst & MBE ($\alpha=2$) \\
 & & & & & CSF ($\tau_\mathrm{SF}=2$\,Myr) & Uniform \\
 & & & & & CSF ($\tau_\mathrm{SF}=2$\,Myr) & MBE ($\alpha=2$) \\
\ref{subsect:dynamic_density_results} (Fig.~\ref{fig:dynamic_density})
    & $10^{-4}$ & 160 & 6.5 & 0.05 & Burst & Uniform (dynamic) \\
\ref{subsect:imf_comparison} (Fig.~\ref{fig:imf_comparison})
    & 0.014 & 160 & 6.0 & 0.05 & Burst & Uniform \\
\ref{subsect:stochastic_results} (Fig.~\ref{fig:stochastic_comparison})
    & 0.014 & 160 & 4.30$^{(a)}$ & 0.05 & Burst & Uniform \\
\hline
\end{tabular}
\tablefoot{\tablefoottext{a}{$M_\mathrm{cl,init} = 2\times10^4\,M_\odot$; stellar mass $M_* = \epsilon_\mathrm{SF}\,M_\mathrm{cl,init} = 10^3\,M_\odot$.}}
\end{table*}

\begin{figure*}
    \centering
    \includegraphics[width=\textwidth]{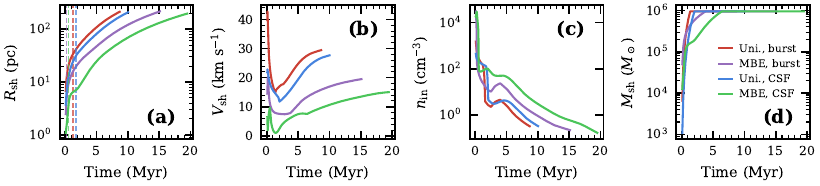}
    \caption{Impact of cloud density profile and star formation mode on feedback-driven shell evolution. Four configurations are compared: uniform density with instantaneous burst (red), uniform density with constant SFR (blue), modified Bonnor-Ebert with instantaneous burst (purple), and modified Bonnor-Ebert with constant SFR (green). The panels show: \textbf{(a)} shell radius, with fragmentation times indicated by dashed vertical lines in the corresponding colors, \textbf{(b)} shell velocity, \textbf{(c)} shell inner edge density, and \textbf{(d)} shell mass. All models share common parameters: $Z = 0.014$, $\epsilon_\mathrm{SF} = 0.05$, $n_\mathrm{cl} = 160$~cm$^{-3}$, $\log\,M_\mathrm{cl,init}/M_\odot = 6.0$. Fragmentation times range from 0.20~Myr (modified Bonnor-Ebert, burst) to 1.73~Myr (uniform, constant SFR).}
    \label{fig:profile_sf_comparison}
\end{figure*}

\subsubsection{Grain size distribution treatment}
\label{subsubsect:grain_size}

We have implemented a more sophisticated treatment of dust grains by modifying the size distribution to better match observations of star-forming regions. The new distribution follows a modified Orion-like profile with an exponential cutoff

\begin{equation}
    \frac{dn}{da} \propto a^{-3.5}\exp\left[-\left(\frac{a_L - a}{\sigma}\right)^n\right],
\end{equation}
where $a$ represents grain size, $a_L = 0.03\,\mu$m is the cutoff size separating small from large grains, $\sigma$ is the width of the exponential cutoff, and $n$ controls the cutoff sharpness, set to 3. The parameter $\sigma$ controls the small-to-large grain mass ratio $r_\mathrm{s/l}$: a small $\sigma$ produces a sharp cutoff that strongly suppresses small-grain mass (Orion-like), while a larger $\sigma$ allows more mass in small grains (ISM-like). For a desired ratio, $\sigma$ is determined numerically by requiring that the integral of $a^3\,dn/da$ over $a < a_L$ divided by the integral over $a > a_L$ matches the target $r_\mathrm{s/l}$. The ratio can be adjusted between $r_\mathrm{s/l} = 0.01$ (Orion-like) and 0.40 (ISM-like), with our default value set to 0.10 for star-forming regions. The upper bound reflects the MRN power law itself: for the adopted size limits, the unmodified $a^{-3.5}$ distribution yields a maximum small-to-large mass ratio of $\sim$0.43, which $\sigma$ can reduce but not exceed. The resulting grain mass distributions for representative ratios are shown in Fig.~\ref{fig:grain_dist}.

The size distribution above applies to the graphite and silicate populations. Polycyclic aromatic hydrocarbons (PAHs) are included separately, with their abundance set by a fixed PAH-to-dust mass fraction (4.6\%, as in the original framework; \citealt{2023MNRAS.526.3871K}) and modulated locally by the atomic hydrogen fraction through \texttt{Cloudy}'s \texttt{grains PAH function} prescription, so that PAHs are suppressed in the ionized gas and survive in the neutral material beyond the ionization front. Each production model is computed through the full swept-up shell (the calculation terminates once the shell mass is reached, well beyond the ionization front), so PAH emission is included in the emergent SEDs.

This parameterization allows for a continuous range of dust properties between the Orion nebula (dominated by large grains) and the diffuse ISM (with more small grains). This two-size description of grain populations is physically well-motivated and consistent with frameworks adopted in galaxy evolution and cosmological dust models \citep{2015MNRAS.447.2937H, 2026MNRAS.545f2040T}, making the \texttt{TODDLERS\,2.0} grain treatment directly relevant to such applications. The effect of the grain size distribution on the emergent SED of an H\,\textsc{ii} region is illustrated in Appendix~\ref{app:grain_sed}.

\begin{figure*}
\centering
\includegraphics[width=0.90\textwidth]{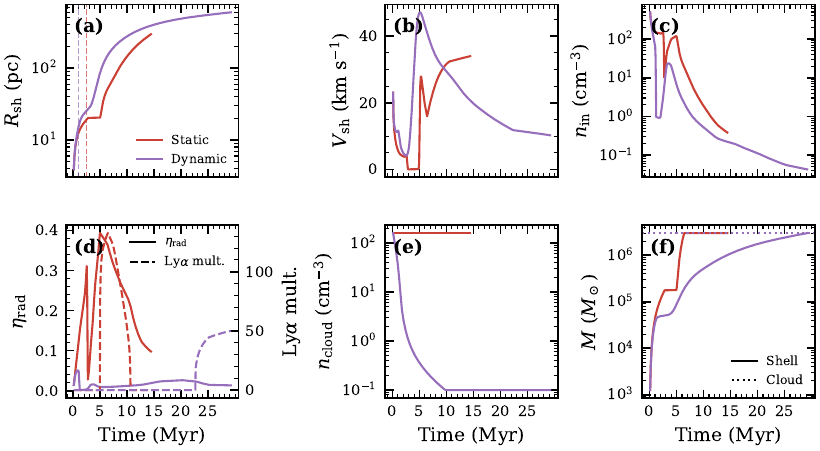}
\caption{Comparison of feedback-driven shell evolution with static (red) and dynamic (purple) cloud density in a very low-metallicity environment ($Z = 10^{-4}$, $n_\mathrm{cl} = 160\,\mathrm{cm}^{-3}$, $\log\,M_\mathrm{cl,init}/M_\odot = 6.5$, $\epsilon_\mathrm{SF} = 0.05$). The panels show: \textbf{(a)} shell radius, \textbf{(b)} shell velocity, \textbf{(c)} shell inner edge density, \textbf{(d)} radiation-to-shell coupling efficiency $\eta_\mathrm{rad}$ (solid; the fraction of bolometric luminosity transferred to the shell) and Lyman-$\alpha$ force multiplier (dashed; the enhancement factor applied to the Lyman-$\alpha$ radiation pressure term), \textbf{(e)} unswept cloud density evolution, and \textbf{(f)} shell mass growth approaching the total cloud mass (dotted lines). Dashed vertical lines in panel~(a) mark the fragmentation times. The unswept cloud density drops by three orders of magnitude over 30~Myr in the dynamic model.}
\label{fig:dynamic_density}
\end{figure*}

\subsubsection{Diffuse ionized gas}
\label{subsubsect:dig}

We include a treatment of diffuse ionized gas (DIG) as a separate \texttt{Cloudy} model that is run after the inner (shell) model. The DIG zone is modeled as a spherical shell at low density ($n_\mathrm{H} = 1\,\mathrm{cm}^{-3}$) starting at the outer radius of the inner model. Three radiation sources illuminate it:

\begin{enumerate}
    \item Direct stellar radiation: a fraction $(1-f_{\mathrm{cover}})$ of the stellar ionizing output reaches the DIG unattenuated, representing photons that escape through gaps in the shell. Here $f_{\mathrm{cover}}$ is the same covering fraction described in Sect.~\ref{subsect:covering_fraction}.
    \item Transmitted radiation: the remaining fraction that intercepts the shell is processed by the inner \texttt{Cloudy} model. The transmitted SED from the inner model is used as an input source, scaled to a luminosity corresponding to $f_{\mathrm{cover}}\, Q_{\mathrm{total}}\, f_{\mathrm{esc,shell}}$, where $f_{\mathrm{esc,shell}}$ is the escape fraction of ionizing photons through the shell.
    \item Old stellar background: a diffuse radiation field from older stellar populations, applied with isotropic illumination geometry using a representative old population SED. Its luminosity is set to produce a user-specified ionization parameter (default $\log U_{\mathrm{bg}} = -3.5$) at the inner radius of the DIG zone.
\end{enumerate}

All three sources are passed to \texttt{Cloudy} simultaneously, which computes the resulting ionization structure self-consistently. Turbulent dissipation heating is included with a default velocity of Mach~0.5 relative to the ionized gas sound speed and a large dissipation length ($\sim$100~kpc) to ensure spatially uniform heating across the DIG zone. After dissolution (Sect.~\ref{subsect:dissolution}), the covering fraction drops to zero and the DIG is illuminated entirely by direct stellar radiation and the old stellar background; the shell-transmitted component is no longer present. The effect of the DIG component on emission-line diagnostics is illustrated in Appendix~\ref{app:bpt_dig}.

Using the \texttt{Cloudy} post-processing infrastructure described above, we have generated pre-computed SED template libraries for use with the radiative transfer code \texttt{SKIRT} \citep{2020A&C....3100381C} in galaxy-scale simulations. The current libraries do not include the DIG component and adopt a fully covering shell ($f_{\mathrm{cover}} = 1$). Templates are available with both a fixed dust-to-metal ratio and with the dust-to-metal ratio as an additional interpolation axis, the latter designed for simulations that track dust masses at the particle level (e.g.\ \texttt{COLIBRE}; \citealt{2025arXiv250821126S}). The available template sets and their parameter ranges are described in Appendix~\ref{app:sed_library}.

\section{Illustrative test cases}
\label{sect:results}

In this section we present a series of test cases that illustrate the impact of the new developments in \texttt{TODDLERS\,2.0}. Each subsection isolates a specific aspect of the framework: the interplay between cloud density profile and star formation mode, the effect of dynamic cloud density evolution in low-metallicity environments, the sensitivity of feedback to IMF variations, the use of custom stellar compositions, and the role of stochastic sampling in low-mass clusters. Sects.~\ref{subsect:profile_sf_comparison} and~\ref{subsect:dynamic_density_results} use the \texttt{BPASS} stellar evolution model with a Chabrier IMF up to 100~$M_\odot$ and binary star evolution; the stellar population choices in Sects.~\ref{subsect:imf_comparison}--\ref{subsect:stochastic_results} are described in the respective subsections. Constant star formation rate models used a formation timescale of $\tau_{\mathrm{SF}} = 2.0$~Myr. Table~\ref{tab:sim_params} summarizes the physical parameters adopted in each subsection.

\begin{figure*}
\centering
\includegraphics[width=0.90\textwidth]{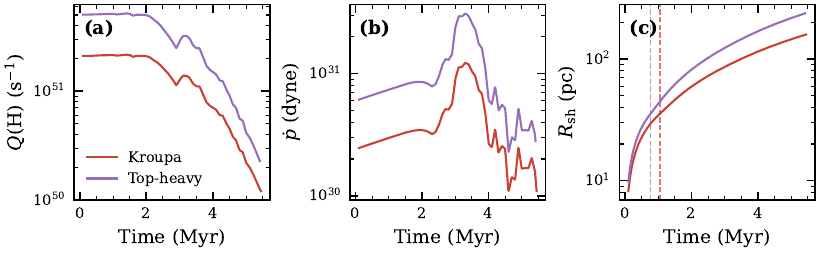}
\caption{Comparison of feedback-driven shell evolution for standard Kroupa (red) and top-heavy (purple) IMFs using \texttt{pySTARBURST99}. The panels show: \textbf{(a)} ionizing photon rate $Q(\mathrm{H})$, \textbf{(b)} ram force, and \textbf{(c)} shell radius evolution. Dashed vertical lines in panel~(c) mark the fragmentation times. Both models use identical cloud parameters: $Z = 0.014$, $\epsilon_{\mathrm{SF}} = 0.05$, $n_{\mathrm{cl}} = 160$~cm$^{-3}$, $\log\,M_{\mathrm{cl,init}}/M_\odot = 6.0$. The top-heavy IMF produces ${\sim}0.5$~dex higher $Q(\mathrm{H})$ and earlier fragmentation (0.77 vs.\ 1.05~Myr).}
\label{fig:imf_comparison}
\end{figure*}

\subsection{Impact of cloud structure and star formation mode on shell evolution}
\label{subsect:profile_sf_comparison}
We illustrate how the cloud density profile and star formation mode affect feedback-driven shell evolution. For this comparison, we use the same set of physical parameters: metallicity $Z = 0.014$ (solar), star formation efficiency $\epsilon_{\mathrm{SF}} = 0.05$, cloud number density $n_{\mathrm{cl}} = 160$ cm$^{-3}$, and initial cloud mass $\log \, M_{\mathrm{cl,init}}/M_\odot = 6.0$. For the modified Bonnor-Ebert profiles (see Fig.~\ref{fig:density_profiles} for the radial density structure), we used a power-law slope of $\alpha = 2$ (equation~\ref{eqn:MBE_equation}) in the outer regions.

Figure~\ref{fig:profile_sf_comparison}, panel~(a) shows the shell radius evolution for each configuration, with fragmentation times indicated in the corresponding colors. The fragmentation times vary significantly across configurations, ranging from 0.20~Myr for the modified Bonnor-Ebert instantaneous case to 1.73~Myr for the uniform constant SFR case. At fixed star formation mode, the modified Bonnor-Ebert profile consistently leads to earlier fragmentation compared to uniform density (0.20~Myr vs. 1.22~Myr for instantaneous bursts, and 0.49~Myr vs. 1.73~Myr for constant SFR). This can be attributed to the steeper density gradient in the modified Bonnor-Ebert profile accelerating the development of instabilities. Additionally, at fixed density profile, constant SFR consistently delays fragmentation compared to instantaneous bursts. This delay occurs because the gradual energy input in constant SFR mode produces more moderate accelerations, reducing the growth rate of Rayleigh-Taylor instabilities. These differences in fragmentation times, and thus the duration of the pressure-driven phase, as well as the energy input rate due to star-formation timescale determine the radial extent of the shells. The modified Bonnor-Ebert constant SFR case shows notably slower expansion due to the combined effects of extended star formation and the higher density concentration in the cloud center.

The velocity evolution in panel~(b) shows distinctive patterns for each configuration. The instantaneous burst models (red and purple) exhibit a sharp initial velocity peak followed by a decline and subsequent re-acceleration after 2-3~Myr. In contrast, the constant SFR models (blue and green) show a more gradual increase in velocity with less pronounced fluctuations, consistent with the more gradual ram force evolution shown in Fig.~\ref{fig:constant_sfr_vs_burst}.

Panel~(c) demonstrates how the inner shell density evolves over time. All configurations show an initial rapid decline followed by fluctuations during fragmentation. The uniform density models maintain slightly lower shell densities in the post-fragmentation phase compared to their modified Bonnor-Ebert counterparts. 

Panel~(d) highlights the shell mass evolution. All configurations eventually sweep up the entire cloud mass, but with different rates of accumulation. The instantaneous burst models accelerate the shell more quickly, leading to faster mass accumulation initially. The constant SFR models show a more gradual approach to the total cloud mass, particularly evident in the modified Bonnor-Ebert constant SFR case (green). This comparative analysis demonstrates that both cloud structure and star formation mode significantly influence the evolution and fragmentation of feedback-driven shells even when the remaining model parameters are the same. The different shell radii across configurations also imply different radiation fluxes at the shell ($\propto L/R^2$), which is expected to affect dust heating and mid-IR emission (e.g.\ at 22\,$\mu$m). Similarly, the differences in shell density (panel c) are expected to influence density-sensitive emission-line ratios such as the [\ion{S}{II}]\,$\lambda\lambda$6716,6731 doublet. These observable consequences will be addressed in the forthcoming parameter study.

\subsection{Dynamic cloud density evolution in low-metallicity environments}
\label{subsect:dynamic_density_results}
As discussed in Sect.~\ref{subsect:dynamic_density}, ionizing radiation escaping through the shell can heat and expand the unswept cloud, reducing the back-pressure on the shell. We choose a very low-metallicity ($Z = 10^{-4}$) massive cloud ($\log \, M_{\mathrm{cl,init}}/M_\odot = 6.5$) for this test case because the dynamic density effect is most pronounced in this regime, as explained below. The initial density is $n_{\mathrm{cl}} = 160\,\mathrm{cm}^{-3}$ with a star formation efficiency of 5\%. Figure~\ref{fig:dynamic_density} shows the results.

The cloud density evolution (panel e) reveals the effect of our newly implemented dynamic density treatment. Over the 30~Myr simulation period, the unswept cloud density decreases by three orders of magnitude, from the initial $160\,\mathrm{cm}^{-3}$ to approximately $0.1\,\mathrm{cm}^{-3}$, approaching the chosen ambient ISM value. This substantial density reduction results from energy injection into the unswept cloud material, primarily through ionizing radiation that escapes the shell. As ionizing photons penetrate beyond the shell, they heat the unswept gas, causing it to expand and reduce in density.

The impact of dynamic cloud density is particularly significant in low-metallicity environments for several reasons. First, stellar populations in metal-poor conditions produce more ionizing photons and have harder spectra, enhancing energy injection into the unswept cloud. Second, the lower dust content in metal-poor gas increases the mean free path of ionizing photons, allowing them to penetrate deeper into the cloud. Third, the weaker stellar winds at low metallicity result in less efficient shell acceleration, making the relative contribution of cloud density evolution to the overall dynamics more pronounced.

This dynamic density implementation affects several aspects of the shell evolution. The shell's velocity profile (panel b) shows a more gradual deceleration after the initial peak around 23\,km\,s$^{-1}$ than would occur in a static-density model. The reduction in cloud density diminishes the external pressure opposing shell expansion, effectively providing an additional feedback channel by modifying the surrounding medium rather than directly pushing the shell. The coupling between radiation and the shell material, quantified by $\eta_{\mathrm{rad}}$ (panel d, solid line), shows complex evolution influenced by the changing cloud density. As the cloud expands and its density decreases, the balance between different feedback mechanisms shifts. Initially, mechanical feedback dominates, but the contribution from radiation pressure, particularly Lyman-$\alpha$ (panel d, dashed line), becomes increasingly important in later stages.

This behavior illustrates a critical insight from our model: in low-metallicity environments, feedback affects not only the swept-up shell but also significantly modifies the ambient medium. Models that maintain static cloud density throughout the evolution may overestimate the deceleration of expanding shells and underestimate their final radii, particularly in metal-poor conditions where radiation effectively couples with the surrounding medium. The simulation also reveals that dynamic cloud density evolution can eventually lead to conditions approaching the ambient ISM, potentially facilitating shell dissolution or breakout into the broader galactic environment. This process may be particularly relevant for understanding feedback effects in dwarf galaxies and high-redshift systems, where low-metallicity conditions prevail.

Having examined how cloud properties and star formation mode shape shell evolution, we now turn to the role of the stellar population itself.

\begin{table}
\caption{Stellar inventories for the three custom populations in Sect.~\ref{subsect:custom_population}.}
\label{tab:custom_populations}
\centering
\begin{tabular}{lccc}
\hline\hline
Population & Stellar inventory & $N_\mathrm{stars}$ & $M_\mathrm{tot}$\,[$M_\odot$] \\
\hline
Massive-dom. & $1{\times}100$, $1{\times}30$, $2{\times}15\,M_\odot$ & 4  & 160 \\
Intermediate & $1{\times}30$, $2{\times}15\,M_\odot$                 & 3  & 60  \\
Low-mass     & $10{\times}8$, $20{\times}5\,M_\odot$                 & 30 & 180 \\
\hline
\end{tabular}
\tablefoot{The massive-dominated population is dominated by a single very massive star; the low-mass population contains no O-type stars ($M > 16\,M_\odot$).}
\end{table}

\begin{figure*}
\centering
\includegraphics[width=0.65\textwidth]{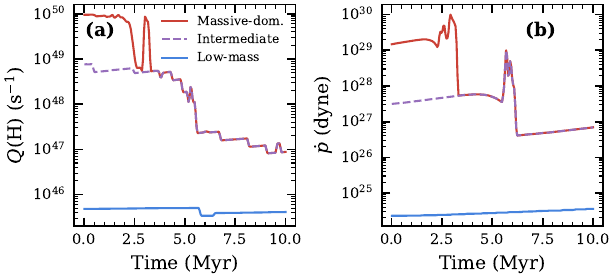}
\caption{Feedback evolution for three custom stellar populations: Massive-dominated (1$\times$100, 1$\times$30, 2$\times$15 $M_\odot$; red, solid), Intermediate (1$\times$30, 2$\times$15 $M_\odot$; purple, dashed), and Low-mass (10$\times$8, 20$\times$5 $M_\odot$; blue, solid). Panel \textbf{(a)} shows the ionizing photon rate $Q(\mathrm{H})$, and panel \textbf{(b)} shows the ram force $\dot{p}$. The massive-dominated population produces ${\sim}1$~dex higher $Q(\mathrm{H})$ and dominates the early wind ram force owing to the $100\,M_\odot$ star, despite only ${\sim}2.7{\times}$ the total stellar mass of the intermediate case. The low-mass population, lacking O-type stars, produces ${\gtrsim}2$~dex lower $Q(\mathrm{H})$ and ram force.}
\label{fig:custom_feedback_examples}
\end{figure*}

\subsection{Impact of the initial mass function on stellar feedback}
\label{subsect:imf_comparison}

The integration of \texttt{pySTARBURST99} into the \texttt{TODDLERS\,2.0} framework enables systematic exploration of how the IMF affects feedback-driven shell evolution. To demonstrate this capability, we compare the evolution of star-forming regions with identical cloud parameters but different IMF prescriptions: a standard Kroupa IMF \citep{2001MNRAS.322..231K} with an upper mass limit of 100~$M_\odot$, and a top-heavy variant with the same mass limits but a shallower high-mass slope ($\alpha_2 = 2.0$ for $M > 0.5\,M_\odot$ instead of the canonical $\alpha_2 = 2.3$).

For this comparison, we adopt fixed cloud parameters: metallicity $Z = 0.014$ (solar), star formation efficiency $\epsilon_{\mathrm{SF}} = 0.05$, cloud number density $n_{\mathrm{cl}} = 160$~cm$^{-3}$, and initial cloud mass $\log\,M_{\mathrm{cl,init}}/M_\odot = 6.0$. Both models use the instantaneous burst mode to isolate the effects of the IMF from star formation history variations.

Figure~\ref{fig:imf_comparison} presents the evolution of key feedback quantities for both IMF prescriptions. Panel~(a) shows the ionizing photon rate $Q(\mathrm{H})$, which directly determines the ionization state of the surrounding gas. The top-heavy IMF produces consistently higher ionizing photon rates, with values approximately 0.5~dex above the standard Kroupa case during the first $\sim$3~Myr. This enhancement arises from the increased fraction of massive O-type stars in the shallower power-law tail, which dominate the ionizing photon budget.

Panel~(b) displays the ram force evolution, which governs the mechanical acceleration of the shell. The top-heavy IMF generates stronger ram forces throughout the evolution, with differences most pronounced during the wind-dominated phase ($t < 3$~Myr). After the onset of supernovae ($t \gtrsim 3$~Myr), the differences diminish as core-collapse events from stars in the $8$--$25\,M_\odot$ range contribute similarly in both models.

Panel~(c) shows the resulting shell radius evolution. The enhanced feedback from the top-heavy IMF leads to faster shell expansion and earlier fragmentation (0.77~Myr vs. 1.05~Myr for Kroupa), reflecting the stronger acceleration-driven instabilities induced by the more vigorous feedback.

These results demonstrate that IMF variations have substantial effects on both the dynamical evolution and observable properties of star-forming regions. The \texttt{pySTARBURST99} integration provides the flexibility to explore arbitrary IMF shapes, enabling investigation of scenarios ranging from bottom-heavy IMFs in metal-rich environments \citep{2010ARA&A..48..339B} to extremely top-heavy distributions in high-redshift starbursts and metal-poor systems \citep{2022ApJ...931...57S, 2024ApJ...973L..32V}.

\subsection{Custom stellar population modeling}
\label{subsect:custom_population}

Beyond IMF-sampled populations, \texttt{pySTARBURST99} enables specification of exact stellar compositions by defining the number of stars at each mass. This capability is valuable for modeling individual stellar associations or for isolating the feedback contribution of specific massive star configurations. Figure~\ref{fig:custom_feedback_examples} compares the feedback output of three custom populations whose stellar inventories are listed in Table~\ref{tab:custom_populations}.

Panel~(a) shows the ionizing photon rate. The massive-dominated population (red) produces an ionizing flux nearly an order of magnitude higher than the intermediate case (purple), despite having only $\sim$2.7 times the mass. The low-mass population (blue) produces negligible ionizing radiation in comparison, illustrating the critical role of massive stars in setting the ionization budget.

Panel~(b) displays the mechanical feedback (winds + supernovae). The presence of the $100\,M_\odot$ star in the massive-dominated case drives strong stellar winds at early times ($t < 3$~Myr). The intermediate and low-mass populations show significantly weaker wind feedback but exhibit distinct features at later times as their specific stellar populations evolve off the main sequence and explode as supernovae.

\subsection{Stochastic stellar population sampling}
\label{subsect:stochastic_results}

\begin{table*}
\caption{Properties of the two stochastic realizations and the deterministic reference shown in Fig.~\ref{fig:stochastic_comparison}. All models have total stellar mass $M_* = 10^3\,M_\odot$.}
\label{tab:stochastic_summary}
\centering
\begin{tabular}{lcccccc}
\hline\hline
Case & Seed & $m_\mathrm{max}$\,[$M_\odot$] & $N_\mathrm{stars}$ & $N(M{>}20\,M_\odot)$ & $t_\mathrm{frag}$\,[Myr] & $R_\mathrm{max}$\,[pc] \\
\hline
Stochastic 1  & 17 & ${\approx}85$ & 352 & 3 & 0.32 & ${\approx}120$ \\
Stochastic 2  & 31 & ${\approx}23$ & 450 & 1 & 1.34 & ${\approx}21$  \\
Deterministic & —  & —             & —   & — & 0.46 & ${\approx}65$  \\
\hline
\end{tabular}
\end{table*}

\begin{figure*}
\centering
\includegraphics[width=0.9\textwidth]{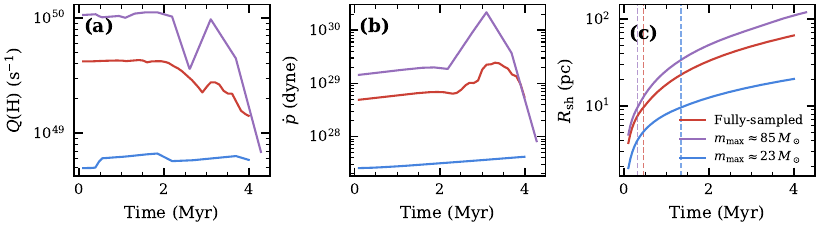}
\caption{Comparison of feedback-driven shell evolution for a fully-sampled Kroupa IMF (red) and two contrasting stochastic realizations with $m_\mathrm{max} \approx 85\,M_\odot$ (purple) and $m_\mathrm{max} \approx 23\,M_\odot$ (blue), all with total stellar mass $M_* = 10^3\,M_\odot$. The two stochastic cases were selected from 50 independent realizations based on the maximum stellar mass drawn. The panels show: \textbf{(a)} ionizing photon rate $Q(\mathrm{H})$, \textbf{(b)} ram force, and \textbf{(c)} shell radius evolution. Dashed vertical lines in panel~(c) mark the fragmentation times. Cloud parameters: $Z = 0.014$, $\epsilon_\mathrm{SF} = 0.05$, $n_\mathrm{cl} = 160\,\mathrm{cm}^{-3}$, $\log\,M_\mathrm{cl,init}/M_\odot = 4.30$. The two stochastic cases differ by ${\sim}1.2$~dex in peak $Q(\mathrm{H})$, ${\sim}2.7$~dex in peak ram force, and a factor of ${\sim}6$ in maximum shell radius.}
\label{fig:stochastic_comparison}
\end{figure*}

The stochastic sampling framework described in Sect.~\ref{subsect:stochastic_sampling} enables modeling of low-mass stellar populations where discrete sampling of the IMF introduces significant variance in the integrated feedback properties. To demonstrate this capability, we survey 50 independent stochastic realizations of a Kroupa IMF population with total stellar mass $M_* = 10^3\,M_\odot$ and an upper mass limit of $100\,M_\odot$ (matching the deterministic comparison), recording the most massive star and number of stars above key mass thresholds. From this survey, we select two contrasting cases based on the maximum stellar mass drawn and compare their feedback-driven evolution against a fully-sampled (deterministic) Kroupa population with identical cloud parameters.

At the mass scale of $10^3\,M_\odot$, the number of massive stars ($M > 20\,M_\odot$) drawn from a Kroupa IMF is of order unity, and the maximum stellar mass varies strongly between realizations. The two selected cases, seed~17 ($m_\mathrm{max} \approx 85\,M_\odot$, 352 stars, 3 stars above $20\,M_\odot$) and seed~31 ($m_\mathrm{max} \approx 23\,M_\odot$, 450 stars, 1 star above $20\,M_\odot$), contain similar total masses ($\sim 10^3\,M_\odot$), but differ drastically in the mass of the most massive member; their properties are listed in Table~\ref{tab:stochastic_summary}. The cloud parameters are: metallicity $Z = 0.014$ (solar), star formation efficiency $\epsilon_\mathrm{SF} = 0.05$, cloud number density $n_\mathrm{cl} = 160\,\mathrm{cm}^{-3}$, and initial cloud mass $\log\,M_\mathrm{cl,init}/M_\odot = 4.30$. All models use the instantaneous burst mode.

Figure~\ref{fig:stochastic_comparison} presents the comparison. Panel~(a) shows the ionizing photon rate $Q(\mathrm{H})$. The $m_\mathrm{max} \approx 85\,M_\odot$ realization produces a peak $Q(\mathrm{H}) \approx 1.1 \times 10^{50}\,\mathrm{s}^{-1}$, which exceeds the fully-sampled deterministic prediction ($4.3 \times 10^{49}\,\mathrm{s}^{-1}$). This occurs because the fully-sampled IMF distributes the $10^3\,M_\odot$ mass budget smoothly, predicting only fractional numbers of stars above $\sim 30\,M_\odot$, whereas the stochastic draw can place a disproportionate share of the mass in a single very massive star. The $m_\mathrm{max} \approx 23\,M_\odot$ case peaks at only $6.7 \times 10^{48}\,\mathrm{s}^{-1}$, over 1.2~dex below the other stochastic realization, as it lacks any contribution from very massive stars, not even the fractional contribution that the fully-sampled IMF predicts.

Panel~(b) shows the ram force $\dot{p}$, where the contrast is even more pronounced: 2.7~dex separates the two stochastic cases ($2.2 \times 10^{30}$ vs.\ $4.2 \times 10^{27}\,\mathrm{dyne}$). The mechanical feedback is dominated by stellar winds from the most massive stars, making it exquisitely sensitive to the high-mass tail of the stochastic draw. The $m_\mathrm{max} \approx 85\,M_\odot$ case also exceeds the deterministic ram force ($2.5 \times 10^{29}\,\mathrm{dyne}$) by nearly an order of magnitude.

Panel~(c) reveals how these differences propagate into shell dynamics. The $m_\mathrm{max} \approx 85\,M_\odot$ realization fragments at $0.32\,\mathrm{Myr}$ and reaches a maximum radius of $\sim$120~pc, while the deterministic case fragments slightly later at $0.46\,\mathrm{Myr}$ and expands to $\sim$65~pc. The $m_\mathrm{max} \approx 23\,M_\odot$ case, lacking vigorous early feedback, fragments much later at $1.34\,\mathrm{Myr}$ and expands to only $\sim$21~pc, a factor of $\sim$6 smaller than the high-$m_\mathrm{max}$ case. That the high-$m_\mathrm{max}$ stochastic realization exceeds the deterministic prediction highlights how, at low cluster masses, the fully-sampled IMF assumption underestimates the feedback from realizations that happen to draw one or more very massive stars.

These results demonstrate the importance of accounting for stochastic sampling effects when modeling feedback in low-mass stellar populations ($M_* \lesssim 10^4\,M_\odot$), where the assumption of a fully-sampled IMF can introduce systematic biases in predicted shell dynamics and observable properties. The stochastic framework implemented here provides a natural way to capture this variance and propagate it through the full feedback-driven evolution.
\section{Summary and outlook}
\label{sec:summary}

We have presented \texttt{TODDLERS\,2.0}, a substantially extended version of the \texttt{TODDLERS} framework for modeling feedback-driven evolution and multi-wavelength emission from star-forming regions (Table~\ref{tab:new_features}). We summarize the main developments and the physical insights gained from our illustrative test cases below.

The framework now interfaces with both \texttt{BPASS} binary evolution (Sect.~\ref{subsubsect:binary_evolution}) and \texttt{pySTARBURST99} (Sect.~\ref{subsect:pysb99}), enabling arbitrary IMF specification, extended mass ranges (up to $300\,M_\odot$ for \texttt{BPASS}, $500\,M_\odot$ for \texttt{pySTARBURST99}), and custom stellar population definitions. A top-heavy IMF yields stronger feedback and earlier fragmentation compared to a standard Kroupa IMF (Fig.~\ref{fig:imf_comparison}), while custom populations allow direct modeling of individual associations (Fig.~\ref{fig:custom_feedback_examples}). For low-mass clusters ($M_* \lesssim 10^4\,M_\odot$), stochastic sampling of individual stellar masses (Sect.~\ref{subsect:stochastic_sampling}) introduces order-of-magnitude variance in feedback properties, confirming the limitations of the fully-sampled IMF assumption in this regime (Fig.~\ref{fig:stochastic_comparison}).

A constant SFR mode (Sect.~\ref{subsect:constant_sfr}) and support for arbitrary density profiles (Bonnor-Ebert, modified Bonnor-Ebert, Gaussian; Sect.~\ref{subsect:density_profiles}) jointly determine the fragmentation timescale and shell extent: centrally concentrated profiles fragment earlier than uniform profiles, and constant SFR consistently delays fragmentation relative to instantaneous bursts (Fig.~\ref{fig:profile_sf_comparison}). Dynamic cloud density evolution (Sect.~\ref{subsect:dynamic_density}) models the heating and expansion of the unswept cloud by escaping ionizing radiation; this effect is most consequential at low metallicities, where the cloud density can drop by three orders of magnitude within $\sim$30~Myr, providing an additional feedback channel missed by static-density models (Fig.~\ref{fig:dynamic_density}).

The framework includes additional refinements to the fragmentation and dissolution criteria, configurable force terms, and a variable ionized gas temperature model in the shell dynamics, as well as modified grain size distributions and diffuse ionized gas modeling in the \texttt{Cloudy} post-processing. While the illustrative test cases focus on features that most directly affect shell dynamics, a comprehensive parameter study exploring the full feature set and its impact on multi-wavelength observables, including emission-line diagnostics and IR dust emission, is planned for a forthcoming publication. The correctness of the framework, through the analytic limits of the shell dynamics and the reduction of each new feature to its validated limit, and its computational performance are presented in Appendices~\ref{app:code_verification} and~\ref{app:performance}.

The framework is being adopted as a sub-grid emission model in radiative transfer post-processing of cosmological simulations \citep{2024A&A...692A..79K, 2026arXiv260105916I}, including the \texttt{COLIBRE}-\texttt{SKIRT} pipeline (Gebek et al., in prep.) and the TNG50-\texttt{SKIRT} Atlas (Baes et al., submitted). The new capabilities extend what is possible in this context: IMF flexibility for environments where a standard IMF may not apply \citep{2022ApJ...931...57S, 2024ApJ...973L..32V}, non-uniform density profiles and dynamic cloud density for more physical cloud descriptions, stochastic sampling for low-mass clusters, and SED templates with a variable dust-to-metal ratio for consistency with simulations that track dust evolution at the particle level (Appendix~\ref{app:sed_library}). In parallel, we have developed a photoionization module for the 3D Monte Carlo radiative transfer code \texttt{SKIRT} that can operate either in tandem with \texttt{TODDLERS} sub-grid models or as a standalone ionized gas treatment (Kapoor et al., submitted).

Independent of its use in galaxy-scale simulations, the predicted dust and PAH emission evolution has been validated against empirical IR SED templates of star clusters observed with \texttt{HST} and \texttt{JWST} in NGC~628 \citep{2025ApJ...982...50W}. The flexible IMF and stochastic sampling extend the range of stellar populations and cluster masses for which such comparisons can be made, and are particularly relevant for star-forming dwarf galaxies, where low cluster masses make the fully-sampled IMF assumption problematic and the resulting variance in ionizing output can affect emission-line ratios and metallicity calibrations \citep{2017MNRAS.470.1612P}. More broadly, the increased flexibility in stellar population and cloud physics makes \texttt{TODDLERS\,2.0} well-suited for interpreting the emission-line diagnostics of H\,\textsc{ii} regions now being delivered by \texttt{JWST} spectroscopic surveys across a wide redshift range.

\section{Data Availability}
The \texttt{TODDLERS\,2.0} source code is publicly available at \url{https://github.com/anandutsavkapoor/toddlers-public}. The repository includes the pre-computed \texttt{pySTARBURST99} stellar evolution data and the \texttt{BPASS} data files required for population synthesis, ensuring that the framework is fully functional out of the box. A working installation of \texttt{Cloudy} is required for the radiative transfer post-processing.

\begin{acknowledgements}
AUK acknowledges support from the Belgian Federal Science Policy Office (BELSPO) via the PRODEX Programme of the European Space Agency (ESA) under contract number 4000143202. SDR acknowledges funding from the FWO. We thank Sarah Sadavoy for discussions that motivated some of the functionality in the code.
\end{acknowledgements}


\FloatBarrier
\bibliographystyle{aa}
\bibliography{bibliography}

\FloatBarrier
\appendix

\section{Configurable force terms}
\label{app:force_terms}

The shell acceleration is determined by
\begin{equation}
    M_{\mathrm{sh}} \frac{dV}{dt} = F_{\mathrm{th}} + F_{\mathrm{rad}} + F_{\mathrm{ram}} - F_{\mathrm{grav}} - F_{\mathrm{ext}} - V_{\mathrm{sh}} \dot{M}_{\mathrm{sh}},
\end{equation}
where $F_{\mathrm{th}}$ is the hot bubble thermal pressure force, $F_{\mathrm{rad}}$ is the radiation pressure, $F_{\mathrm{ram}} = \dot{p}$ is the ram force from mass loss (stellar winds and supernovae), $F_{\mathrm{grav}}$ is the gravitational force, $F_{\mathrm{ext}}$ is the external confining pressure, and the final term accounts for momentum loading from swept-up mass. Of these, the gravitational, Lyman-$\alpha$ radiation pressure, and external pressure terms can be selectively enabled. Their relative significance depends on cloud mass, metallicity, and evolutionary stage; selectively disabling terms allows users to isolate individual contributions and reproduce the assumptions of simpler analytical models for comparison.

The gravitational force accounts for the combined gravitational pull of the central stellar cluster, the shell self-gravity (approximated as half the shell mass), and any enclosed mass
\begin{equation}
    F_{\mathrm{grav}} = \frac{G \, M_{\mathrm{sh}} \left( M_* + \tfrac{1}{2}M_{\mathrm{sh}} \right)}{R^2}.
\end{equation}
The factor $\tfrac{1}{2}$ is the standard self-gravity coefficient for a thin shell, which experiences the mean of the vanishing interior field and the full exterior field; since the swept-up shell stays geometrically thin, resolving its finite radial extent yields a negligible correction. This term is enabled by default. The radiation force includes a standard UV$+$IR component $F_{\mathrm{rad,UV+IR}} = \eta_{\mathrm{rad}} \, L_{\mathrm{bol}} / c$ that is always computed, plus a Lyman-$\alpha$ resonant scattering term following the formalism of \citet{2018MNRAS.475.4617K}. The Lyman-$\alpha$ force is computed as
\begin{equation}
    F_{\mathrm{Ly}\alpha} = M_{\mathrm{boost}} \, \frac{L_{\mathrm{Ly}\alpha,\mathrm{dusty}}}{c},
\end{equation}
where $L_{\mathrm{Ly}\alpha,\mathrm{dusty}}$ is the Lyman-$\alpha$ luminosity attenuated by dust within the ionized shell, and $M_{\mathrm{boost}}$ is a force multiplier that depends on the neutral hydrogen column density, dust optical depth, and shell velocity. The multiplier is reduced at high shell velocities ($V_{\mathrm{sh}} \gtrsim 100$~km~s$^{-1}$) where Doppler shifts break the resonance condition. This term is enabled by default.

The external pressure from the ambient medium, either the unswept cloud or the ISM after the cloud has been fully swept, exerts a confining force on the shell. When ionizing photons escape the shell into the unswept cloud, the external pressure transitions from the neutral cloud thermal pressure to the higher ionized pressure $P_{\mathrm{ion}} = P_{\mathrm{neutral}} (\mu_N / \mu_P)(T_{\mathrm{ion}} / T_{\mathrm{neutral}})$, using a smooth interpolation based on the ionizing escape fraction. This term is enabled by default.

\section{Variable ionized gas temperature}
\label{app:variable_tion}

The interpolator maps metallicity $Z$, hydrogen density $n_{\mathrm{H}}$, and ionizing photon flux $\phi$ to the equilibrium temperature $T_{\mathrm{ion}}(Z, n_{\mathrm{H}}, \phi)$, where $\phi = Q(\mathrm{H}) / (4\pi R^2)$ is the local ionizing flux at the shell radius. The grid models adopt a fixed \texttt{BPASS} stellar population spectrum at an age of 1\,Myr as the ionizing source, spanning five metallicities ($Z = 0.001$--$0.04$), ten hydrogen densities ($n_\mathrm{H} = 10^{-3}$--$10^{6}$\,cm$^{-3}$), and eleven ionizing flux values covering the parameter space relevant to the cluster masses and radii considered here. We note that the equilibrium temperature of photoionized gas is set primarily by the balance between photoionization heating and metal-line cooling, and is therefore most sensitive to metallicity. The spectral shape of the ionizing source enters only through the mean photon energy, which varies modestly between 1 and 3~Myr for a \texttt{BPASS} population. The 1\,Myr spectrum is representative of the age range during which most of the ionizing luminosity is produced.

At each timestep, the shell inner-face density and temperature are determined simultaneously by solving the implicit equation
\begin{equation}
    n = \frac{\mu_P \, P}{\mu_N \, k_{\mathrm{B}} \, T_{\mathrm{ion}}(Z, n, \phi)},
\end{equation}
where $P$ is the pressure at the shell's inner face. The equilibrium temperature is governed primarily by the metallicity, which sets the abundance of line coolants, and by the ionizing photon flux $\phi$, which drives photoionization heating. Density enters only secondarily: at very high densities, collisional de-excitation can quench line emission and reduce cooling efficiency. At fixed pressure, a higher metallicity leads to more efficient cooling and thus a lower $T_{\mathrm{ion}}$, which in turn requires a higher inner-face density to maintain pressure balance. The self-consistent temperature obtained from this implicit solution propagates into the shell structure equations and the sound speed used in fragmentation criteria.

\section{Effect of grain size distribution on the SED}
\label{app:grain_sed}

\begin{figure}
\centering
\includegraphics[width=0.85\linewidth]{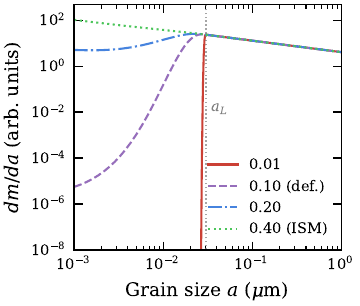}
\caption{Grain mass distribution $dm/da$ as a function of grain size for four small-to-large grain mass ratios $r_\mathrm{s/l}$. The dotted line marks the cutoff size $a_L = 0.03\,\mu$m.}
\label{fig:grain_dist}
\end{figure}

Figure~\ref{fig:grain_dist} shows the grain mass distribution $dm/da$ for four representative values of the small-to-large grain mass ratio $r_\mathrm{s/l}$. The cutoff at $a_L = 0.03\,\mu$m (dotted line) separates the small and large grain populations. Lower ratios produce a sharper suppression of small grains below $a_L$, while higher ratios approach the unmodified MRN power law.

\begin{figure}
\centering
\includegraphics[width=0.85\linewidth]{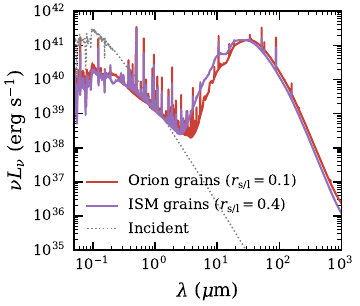}
\caption{Comparison of emergent SEDs from a \texttt{Cloudy} H\,\textsc{ii} region model with Orion-like (small-to-large grain ratio = 0.10, red) and ISM-like (ratio = 0.40, purple) grain size distributions. The incident spectrum is shown in gray. The model covers the ionized zone only.}
\label{fig:grain_sed}
\end{figure}

Figure~\ref{fig:grain_sed} compares the emergent SED from a single-zone \texttt{Cloudy} H\,\textsc{ii} region model computed with Orion-like (small-to-large ratio = 0.10) and ISM-like (ratio = 0.40) grain size distributions. Both models use the same ionizing source (a 1\,Myr \texttt{BPASS} Chab100 binary population with $\log Q(\mathrm{H}) = 51.5$), hydrogen density $n_\mathrm{H} = 100\,\mathrm{cm}^{-3}$, and solar metallicity. To isolate the effect of the graphite and silicate grain size distribution on the continuum, this test model is restricted to the ionized zone (stopping at an electron fraction of 0.1); PAH emission features are therefore absent from this particular figure, since PAHs survive only in the neutral material beyond the ionization front. In the full production models, which extend through the entire swept-up shell, PAH emission is included (Sect.~\ref{subsubsect:grain_size}). The ISM-like distribution contains a larger fraction of small grains, which absorb more efficiently in the UV and re-emit more strongly in the MIR through stochastic heating.

\section{Effect of diffuse ionized gas on BPT diagnostics}
\label{app:bpt_dig}
\begin{figure}
\centering
\includegraphics[width=0.85\linewidth]{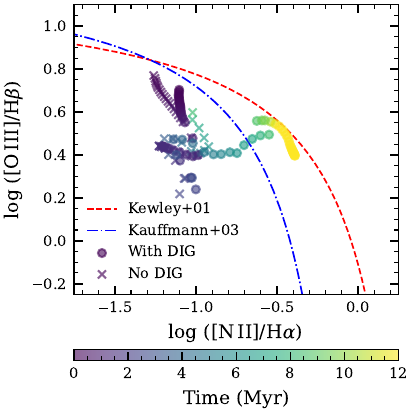}
\caption{BPT diagram showing the effect of diffuse ionized gas on emission-line ratios. Circles: shell + DIG; crosses: shell only. Color indicates evolutionary time. The \citet{2001ApJ...556..121K} and \citet{2003MNRAS.346.1055K} demarcation curves are shown for reference.}
\label{fig:bpt_dig}
\end{figure}
Figure~\ref{fig:bpt_dig} shows the effect of including the DIG component on emission-line ratios in the {\NII}/H$\alpha$ vs.\ {\OIII}/H$\beta$ BPT diagram. Both cases use the same evolution model ($Z = 0.014$, $n_\mathrm{cl} = 160\,\mathrm{cm}^{-3}$, $\log\,M_\mathrm{cl,init}/M_\odot = 6.0$, $\epsilon_\mathrm{SF} = 0.05$, \texttt{BPASS} Chab100 binary, burst mode) with full covering ($f_\mathrm{cover} = 1$). In this configuration, the DIG is illuminated by radiation transmitted through the shell and by an isotropic old stellar background ($\log U_\mathrm{bg} = -3.5$). As the shell thins, an increasing fraction of ionizing photons reaches the DIG directly. At early times, when the shell is optically thick and dust absorption is significant, the DIG contribution shifts the emergent ratios (circles) rightward relative to the shell-only ratios (crosses). The \texttt{BPASS} binary populations produce harder spectra at late times, which, together with the old stellar background heating, place the DIG emission between the \citet{2001ApJ...556..121K} and \citet{2003MNRAS.346.1055K} demarcation lines. After the shell dissolves, only the DIG component remains and the two sets of points converge. Points are color-coded by evolutionary time.

\section{SED template library}
\label{app:sed_library}

The \texttt{TODDLERS} SED templates are publicly available as \texttt{SKIRT} resource packs\footnote{\url{https://skirt.ugent.be/skirt9/class_toddlers_s_e_d_family.html}} and are distributed as stored tables (\texttt{.stab} files) that can be used with the \texttt{ToddlersSEDFamily} class in \texttt{SKIRT~9}. Two modes are available: Cloud mode, which provides time-resolved SEDs for individual clouds (for a given $M_\mathrm{cl}$), and SFRNormalized mode, which provides SEDs pre-integrated over the cloud mass function and time, directly scaled by star formation rate.

\subsection{Available template sets}

Table~\ref{tab:stab_library} summarizes all available SED template sets. Three generations of templates are currently distributed with \texttt{SKIRT}. The v1 templates were generated with the original \texttt{TODDLERS} framework presented in \citet{2023MNRAS.526.3871K}, using \texttt{STARBURST99} single-star populations with a Kroupa IMF. The v2 templates use \texttt{BPASS} binary populations with a Chabrier IMF and were generated with the version of the code presented in this paper; they are distributed with \texttt{SKIRT} and this paper provides their first description in the literature. The v2-DTM templates extend the v2 set by including a dust scaling factor $f_\mathrm{dust}$ as an additional interpolation axis (Sect.~\ref{app:dtm_calibration}), allowing the subgrid emission model to reflect the resolved dust content of each resolution element. Both v1 and v2 adopt a fixed dust content ($f_\mathrm{dust} = 1$) and are suitable for simulations that do not explicitly track dust.

\begin{table*}[h]
\centering
\caption{Summary of available \texttt{TODDLERS} SED template sets. All templates are distributed as \texttt{SKIRT} stored tables.}
\label{tab:stab_library}
\begin{tabular}{llllcc}
\hline
Version & Mode & Stellar template & Axes & $f_\mathrm{dust}$ & Small-to-large ratio \\
\hline
v1 & SFRNormalized & \texttt{SB99} Kroupa100 & $\lambda$, $Z$, $\epsilon_\mathrm{SF}$, $n_\mathrm{cl}$ & 1.00$^{*}$ & Orion$^{*}$ \\
v1 & Cloud & \texttt{SB99} Kroupa100 & $\lambda$, $t$, $Z$, $\epsilon_\mathrm{SF}$, $n_\mathrm{cl}$, $M_\mathrm{cl}$ & 1.00$^{*}$ & Orion$^{*}$ \\
v2 & SFRNormalized & \texttt{BPASS} Chab100 & $\lambda$, $Z$, $\epsilon_\mathrm{SF}$, $n_\mathrm{cl}$ & 1.00 & 0.10 \\
v2 & Cloud & \texttt{BPASS} Chab100 & $\lambda$, $t$, $Z$, $\epsilon_\mathrm{SF}$, $n_\mathrm{cl}$, $M_\mathrm{cl}$ & 1.00 & 0.10 \\
v2-DTM & SFRNormalized & \texttt{BPASS} Chab100 & $\lambda$, $Z$, $\epsilon_\mathrm{SF}$, $n_\mathrm{cl}$, $f_\mathrm{dust}$ & 0.02--1.00 & 0.40 \\
\hline
\end{tabular}
\tablefoot{$f_\mathrm{dust}$ is the grain abundance scaling factor (Sect.~\ref{app:dtm_calibration}); $f_\mathrm{dust} = 1$ corresponds to $\mathrm{DTM} = 0.456$ (v2) or 0.39 (v1) at solar metallicity. $^{*}$The v1 templates use \texttt{Cloudy}'s built-in \texttt{grains Orion} command; the grain size distribution was not independently configurable in that version.}
\end{table*}

\begin{table*}[h]
\centering
\caption{SED components included in each template configuration. ``low'' and ``high'' refer to the spectral resolution: low uses $R = 300$ throughout, high adds emission lines at $R = 5 \times 10^4$ on top of the low-resolution continuum. See the \texttt{SKIRT} documentation for details on the line replacement procedure.}
\label{tab:sed_components}
\begin{tabular}{l cc cc}
\toprule
 & \multicolumn{2}{c}{\texttt{includeDust=true}} & \multicolumn{2}{c}{\texttt{includeDust=false}} \\
\cmidrule(lr){2-3} \cmidrule(lr){4-5}
Component & low & high & low & high \\
\midrule
\multicolumn{5}{c}{\textbf{v1 and v2}} \\
\addlinespace
Stellar continuum     & attenuated & attenuated & incident     & incident \\
Nebular continuum     & \checkmark & \checkmark & --           & --       \\
Nebular lines         & \checkmark & emergent   & --           & intrinsic \\
Dust thermal emission & \checkmark & \checkmark & --           & --       \\
\addlinespace
\multicolumn{5}{c}{\textbf{v2-DTM}} \\
\addlinespace
Stellar continuum     & attenuated & attenuated & incident     & incident \\
Nebular continuum     & \checkmark & \checkmark & unattenuated & unattenuated \\
Nebular lines         & \checkmark & emergent   & --           & intrinsic \\
Dust thermal emission & \checkmark & \checkmark & --           & --       \\
\bottomrule
\end{tabular}
\end{table*}

\subsection{Variable dust-to-metal ratio templates}
\label{app:dtm_calibration}

\texttt{Cloudy} defines grain abundances per hydrogen atom. In our templates, this abundance is scaled by $Z/Z_\odot$, so the dust mass per hydrogen atom is strictly proportional to metallicity. The conventional dust-to-metal ratio normalizes the dust mass by the total gas mass rather than the hydrogen mass alone,
\begin{equation}
    \mathrm{DTM} \equiv \frac{\mathrm{D/G}}{Z} = \frac{m_\mathrm{dust}}{m_\mathrm{gas}} \cdot \frac{1}{Z} = \frac{m_\mathrm{dust}/m_\mathrm{H}}{\mu_\mathrm{gas} \cdot Z}\,,
\end{equation}
where $\mu_\mathrm{gas} = m_\mathrm{gas}/m_\mathrm{H}$ accounts for helium in the total gas mass. Because the helium mass fraction increases with $Z$ through the adopted He/H--$Z$ relation \citep{2023MNRAS.526.3871K}, $\mu_\mathrm{gas}$ grows with metallicity and DTM varies by $\sim$20\% across the metallicity range even though the dust content per hydrogen atom is exactly proportional to $Z$. At solar metallicity ($Z = 0.014$, $\mu_\mathrm{gas} \approx 1.41$), the v2 templates yield $\mathrm{DTM} = 0.456$, corresponding to a dust-to-hydrogen mass ratio per unit metallicity of
\begin{equation}
    \xi \equiv \frac{m_\mathrm{dust}}{m_\mathrm{H}\,Z} = \mathrm{DTM} \cdot \mu_\mathrm{gas} = 0.643\,,
\end{equation}
which is constant across metallicities. The v1 templates yield $\mathrm{DTM} = 0.39$, the difference arising from the 0.85$\times$ scaling factor applied internally by \texttt{Cloudy}'s \texttt{grains Orion} command ($0.85 \times 0.456 \approx 0.39$).

The v2-DTM SFRNormalized templates extend the v2 set by including a dust scaling axis. The axis stores a dimensionless factor $f_\mathrm{dust}$ that directly multiplies the grain abundance per hydrogen atom in the \texttt{Cloudy} input ($f_\mathrm{dust} = 1$ corresponds to the full v2 grain content; the v1 and v2 templates are generated at $f_\mathrm{dust} = 1$). Because it scales a per-hydrogen-atom quantity, $f_\mathrm{dust}$ is metallicity-independent by construction, making it the natural axis choice. The axis spans seven values: $f_\mathrm{dust} = [0.02,\, 0.10,\, 0.20,\, 0.40,\, 0.60,\, 0.80,\, 1.00]$, chosen to cover the range of dust fractions found in star-forming gas across $0 \leq z \leq 2$ in \texttt{COLIBRE} cosmological simulations with a live dust model \citep{2025arXiv250821126S}. The v2-DTM templates are generated using the same evolution simulations as the v2 templates, with both the dust scaling and the grain size ratio (see below) applied only in the \texttt{Cloudy} post-processing step.

For a simulation that tracks dust and hydrogen masses per resolution element, the appropriate $f_\mathrm{dust}$ for template lookup is
\begin{equation}
    f_\mathrm{dust} = \frac{m_\mathrm{dust}/m_\mathrm{H}}{\xi \, Z}\,,
\end{equation}
where $\xi = 0.643$ is the dust-to-hydrogen mass ratio per unit metallicity at full dust content. If the simulation instead tracks D/G, then
\begin{equation}
    f_\mathrm{dust} = \frac{\mathrm{D}}{\mathrm{G}} \cdot \frac{\mu_\mathrm{gas}(Z)}{\xi \, Z}\,,
\end{equation}
where $\mu_\mathrm{gas}(Z) = m_\mathrm{gas}/m_\mathrm{H}$ depends on the assumed He/H--$Z$ relation.

Each template version provides two SED variants, controlled by the \texttt{includeDust} parameter in \texttt{SKIRT}. Table~\ref{tab:sed_components} summarizes the components included in each case. With \texttt{includeDust=true}, the full \texttt{Cloudy} output is used, including the stellar continuum processed through the shell, nebular continuum and line emission, and dust thermal emission. In the high-resolution versions, the emission lines are replaced by high-resolution emergent line profiles, where the emergent luminosities account for foreground attenuation by dust and gas opacity within the cloud. With \texttt{includeDust=false}, the continuum is the unprocessed incident stellar radiation field. In the high-resolution versions, intrinsic emission line luminosities from \texttt{Cloudy} are added; these reflect the ionization structure of the full model (including the effect of dust on the radiation field) but do not include foreground attenuation of the line luminosities themselves. The v2-DTM \texttt{includeDust=false} SEDs additionally include the unattenuated diffuse nebular continuum, which required a modification to the \texttt{Cloudy} output routines to save this component separately. We note that in both variants, dust is present in the \texttt{Cloudy} model and affects the ionization structure and temperature; the difference is whether the output SED includes the dust-reprocessed emission and foreground attenuation. As a result, the nebular emission in the \texttt{includeDust=false} SEDs depends on $f_\mathrm{dust}$ even though no dust attenuation or thermal emission appears in the output.

The small-to-large grain mass ratio (Sect.~\ref{subsubsect:grain_size}) for the v2-DTM templates is set to 0.40 (ISM-like), compared to 0.10 (Orion-like) for the v2 templates, based on the grain species fractions predicted by \texttt{COLIBRE} cosmological simulations at $0 \leq z \leq 2$.

\section{Equivalent Evolutionary Point interpolation}
\label{app:eep}

In the stochastic mode (Sect.~\ref{subsect:mass_assignment}), each star's feedback is read from a single-star database tabulated on a discrete grid of stellar masses. To interpolate between grid masses without averaging stars in unlike evolutionary phases, we align them by Equivalent Evolutionary Point (EEP) before interpolating, following the prescription of \citet{2016ApJS..222....8D}. We adapt the EEP concept to massive-star feedback rather than reproduce the full MIST isochrone framework: we retain only the markers relevant to the feedback-producing phases (main sequence through the Wolf-Rayet stage) and omit the later stages (horizontal branch, asymptotic giant branch, and white dwarf cooling), which neither occur nor contribute on the timescales we model. The primary markers are listed in Table~\ref{tab:eep}. Each is grounded either in a MIST primary-EEP criterion (where it is expressible in our track observables), in a \texttt{pySTARBURST99} spectral-classifier boundary (for the hot and Wolf-Rayet regimes, which have no MIST massive-star analogue), or in a feedback-relevant temperature threshold; MIST's massive-star primary EEPs cannot be used verbatim because they key off quantities our tracks do not carry. Secondary EEPs are distributed between consecutive primaries at equal increments of the MIST Hertzsprung-Russell metric (\citealt{2016ApJS..222....8D}, their Eq.~1). Interpolating at fixed EEP rather than fixed age keeps the bracketing tracks in the same evolutionary phase, so a Wolf-Rayet star is never blended with a main-sequence star of the same age.

\begin{table}[h]
\centering
\caption{Primary EEP markers used for phase-aligned mass interpolation. Source: M, a MIST primary-EEP phase \citep{2016ApJS..222....8D} located from our track observables (the $T_\mathrm{eff}$ minimum is the criterion \citeauthor{2016ApJS..222....8D} adopt for the RGB tip; ZAMS and the turn-off stand in for their central-abundance definitions); P, a \texttt{pySTARBURST99} spectral-classifier boundary; F, a feedback temperature threshold.}
\label{tab:eep}
\footnotesize
\begin{tabular}{@{}l p{3.6cm} l@{}}
\hline
Marker & Definition & Source \\
\hline
ZAMS & track start & M \\
MS turn-off & $\log T_\mathrm{eff}$ falls 0.05 dex below its running maximum & M \\
$T_\mathrm{down}$ & $\log T_\mathrm{eff} < 4.45$ (ionizing-decline onset) & F, P \\
$T_\mathrm{eff,min}$ & coolest point, $\arg\min \log T_\mathrm{eff}$ & M, P \\
WN onset & surface $X(\mathrm{H}) < 0.4$ & P \\
\hline
\end{tabular}
\end{table}

We validate the scheme with a leave-one-out test: each grid mass is removed in turn and reconstructed from its neighbours by EEP interpolation. The reconstructed time-resolved feedback recovers the held-out track to within $\approx 0.05$~dex for the ionizing output and $0.02$ to $0.05$~dex for the winds, at all four metallicities. The deterministic fully-sampled synthesis already interpolates in mass (linearly at fixed evolutionary-row index) and agrees with the EEP result to within $\approx 3$ per cent of the integrated budget; we therefore retain it in the fully-sampled regime and offer EEP as the validated interpolating option in the stochastic per-star path.

\section{Code verification}
\label{app:code_verification}

\texttt{TODDLERS\,2.0} extends the framework introduced and validated in \citetalias{2023MNRAS.526.3871K}: the 1D shell-dynamics core, the \texttt{Cloudy} post-processing chain, and the feedback coupling are inherited from that work and are not re-derived here. \citetalias{2023MNRAS.526.3871K} initialized the shell from the analytic \citet{1977ApJ...218..377W} wind-bubble solution; it found its approximate shell density structure consistent with detailed \texttt{Cloudy} calculations (the mass-weighted density agreeing to within 10\% over a subset of the parameter space), compared the predicted infrared colors against an independent \texttt{MAPPINGS\,III}-based model, and matched the observed colors of Galactic and extragalactic star-forming regions; the shell-dynamics model was also benchmarked directly against \texttt{WARPFIELD} \citep{2017MNRAS.470.4453R, 2019MNRAS.483.2547R} for homogeneous clouds. We do not attempt an exhaustive verification of every code path here. Instead, we provide two targeted checks that the new machinery is physically sound: an explicit analytic-limit verification that the integrator, as run here, reproduces the analytic dynamical limits where they exist, and a confirmation that each new feature reduces to the validated v1 behavior in the appropriate limit.

\subsection{Analytic limits of the shell dynamics}

A feedback-driven shell expanding into a uniform medium passes through self-similar regimes with constant logarithmic slopes $d\ln R/d\ln t$. While thermal pressure from the shocked wind dominates (the energy-driven phase), energy conservation gives $R(t) = \left(250/308\pi\right)^{1/5}(L_\mathrm{w}/\rho_0)^{1/5}\,t^{3/5}$ for a constant mechanical luminosity $L_\mathrm{w}$: a slope of $3/5$ with self-similar constant $0.763$ \citep{1977ApJ...218..377W}. After fragmentation the shell is driven by direct momentum injection while sweeping ambient mass ($M_\mathrm{sh}\propto\rho_0 R^3$); its slope then lies between two limits, $1/4$ for a freely coasting (momentum-conserving) snowplow and $1/2$ for one in which a constant momentum-injection rate dominates.

We drive the full \texttt{TODDLERS\,2.0} integrator with idealized but realistic inputs, with gravity, Lyman-$\alpha$, and external pressure disabled and the UV$+$IR radiation force set to zero (Appendix~\ref{app:force_terms}), so that the wind alone acts. A single wind carries both energy and momentum: for a mechanical luminosity $L_\mathrm{w}=\tfrac12\dot{M}v_\mathrm{w}^2$ and terminal velocity $v_\mathrm{w}$, the momentum-injection rate is $\dot{p}=\dot{M}v_\mathrm{w}=2L_\mathrm{w}/v_\mathrm{w}$. We hold $L_\mathrm{w}$ constant and set $\dot{p}=2L_\mathrm{w}/v_\mathrm{w}$ with $v_\mathrm{w}=2870$~km~s$^{-1}$, the energy-weighted effective velocity $2\langle L_\mathrm{w}\rangle/\langle\dot{p}\rangle$ of the population measured from the \texttt{STARBURST99} tables, so that the same wind sets both phases. The same set of runs, $n_0\in\{10,100,1000\}$~cm$^{-3}$ and $L_\mathrm{w}\in\{10^{38},10^{39}\}$~erg~s$^{-1}$ (bracketing the $\sim$5$\times10^{38}$~erg~s$^{-1}$ of the standard cluster), is shown in both panels. No dynamics-core code is modified.

\begin{figure*}
\centering
\includegraphics[width=0.7\linewidth]{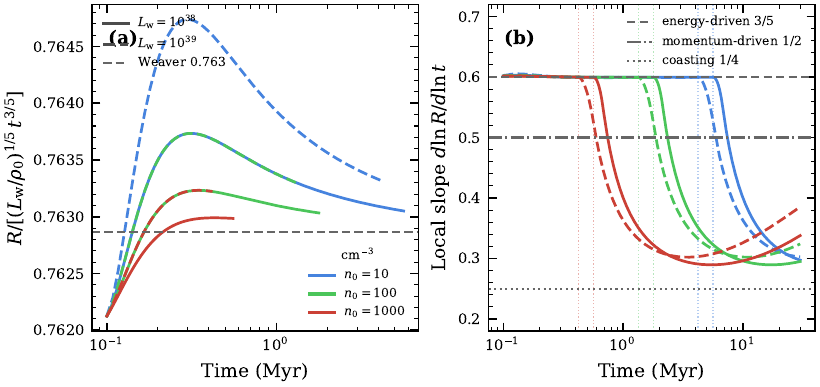}
\caption{Analytic-limit verification of the shell dynamics; the same idealized runs ($n_0\in\{10,100,1000\}$~cm$^{-3}$, colors; $L_\mathrm{w}\in\{10^{38},10^{39}\}$~erg~s$^{-1}$, line styles) appear in both panels. \textit{Left:} in the energy-driven phase the self-similar variable $R/[(L_\mathrm{w}/\rho_0)^{1/5}t^{3/5}]$ settles onto the \citet{1977ApJ...218..377W} constant $0.763$ (dashed) to within $0.3\%$. \textit{Right:} the local logarithmic slope $d\ln R/d\ln t$ plateaus near the energy-driven value $3/5$, then after fragmentation (dotted vertical lines) lies between the coasting ($1/4$) and constant-injection snowplow ($1/2$) limits, trending toward $1/2$; curves are truncated at cloud exhaustion.}
\label{fig:weaver}
\end{figure*}

Figure~\ref{fig:weaver} shows the result. In the energy-driven phase (panel a), the self-similar variable $R/[(L_\mathrm{w}/\rho_0)^{1/5}t^{3/5}]$ settles onto the Weaver constant $(250/308\pi)^{1/5}=0.763$ to within $0.3\%$ across all six runs, confirming both the $t^{3/5}$ scaling and its normalization. In panel (b) the local slope plateaus near $3/5$ until fragmentation, beyond which the bubble can no longer confine its thermal pressure and the shell is driven by the wind and supernova ram force. For constant $\dot{p}$ the momentum equation $d(M_\mathrm{sh}V)/dt = F$ integrates exactly to $R^4 = (3\dot{p}/2\pi\rho_0)\,t^2 + (3p_0/\pi\rho_0)\,t + R_0^4$, with $p_0\equiv(M_\mathrm{sh}V)_\mathrm{frag}$; the injected ($\propto t^2$) and inherited-momentum ($\propto t$) terms drive $R\propto t^{1/2}$ and $t^{1/4}$, so the slope stays between these limits and approaches $1/2$ only near $t\simeq2p_0/\dot{p}$, late for the realistic, sub-dominant $\dot{p}$. Each curve is truncated at cloud exhaustion ($M_\mathrm{sh}\to M_\mathrm{cl}$), beyond which $M_\mathrm{sh}\propto\rho_0 R^3$ fails. Throughout, the driving force is balanced mainly by the swept-mass loading $V\dot{M}_\mathrm{sh}$ rather than the bulk inertia $M_\mathrm{sh}\dot{V}$: the energy-driven phase reproduces the self-similar ratio $F_\mathrm{th}:V\dot{M}_\mathrm{sh}:|M_\mathrm{sh}\dot{V}| = 21:27:6$, and the integrator satisfies $M_\mathrm{sh}\dot{V} + V\dot{M}_\mathrm{sh} = F$ to numerical precision.

\subsection{Reduction of the new features to validated limits}

The analytic-limit checks above, together with the validation of the engine and the shared \texttt{Cloudy} stage in Paper I, establish the correctness of the core model. Each new \texttt{TODDLERS\,2.0} feature is a controlled extension of that core: in its disabled or limiting state it reduces to the validated baseline (the v1 model, or its fully-sampled deterministic limit), so no new feature can silently perturb it. Table~\ref{tab:verification} collects these reductions, which are exercised by the test suite distributed with the code.

\begin{table*}[h]
\centering
\caption{Reduction of the new \texttt{TODDLERS\,2.0} features to the validated baseline. In its disabled or limiting state, each feature recovers the v1 model (or its fully-sampled deterministic limit), confirming that the extensions are self-consistent with the validated core.}
\label{tab:verification}
\begin{tabular}{ll}
\hline
Feature & Verification \\
\hline
Dynamic cloud density & Recovers the static-density model when the cloud density is held fixed. \\
Non-uniform density profiles & Reduce to the uniform case for a flat profile; initial radius from the enclosed-mass root-find. \\
Variable covering fraction & $f_\mathrm{cover}=1$ recovers the fully covered v1 shell; forces scale linearly with $f_\mathrm{cover}$. \\
Fragmentation criteria & Reduce to the v1 pressure ($\beta$) criterion; additional criteria are physically motivated thresholds. \\
Stochastic IMF sampling & Converges to the deterministic fully-sampled integral as mass increases; in the under-sampled \\
 & regime the mass-assignment choice is dominated by realization scatter (Sect.~\ref{subsect:stochastic_sampling}). \\
Variable dust-to-metal ratio & $f_\mathrm{dust}$ scales the grain abundance per H atom and reduces to the baseline v2 grain content at \\
 & $f_\mathrm{dust}=1$; the resulting dust-to-metal ratio is derived in Appendix~\ref{app:dtm_calibration}. \\
Diffuse ionized gas & Optional component; with it disabled the model recovers the v1 shell-only emission \\
 & (Appendix~\ref{app:bpt_dig}). \\
\hline
\end{tabular}
\end{table*}

\section{Computational performance}
\label{app:performance}

\texttt{TODDLERS\,2.0} retains the delivery model of v1: the expensive physics (the 1D dynamics and the time-resolved \texttt{Cloudy} post-processing) is computed once to build a grid, and end users consume precomputed interpolation tables. These tables can be queried directly as standalone interpolants, or loaded into \texttt{SKIRT} as the stored tables of Appendix~\ref{app:sed_library}. The cost therefore separates into three regimes.

The first is table evaluation by the end user. Generating an SED for an arbitrary set of parameters is an interpolation in a precomputed table, a sub-millisecond operation. An inference pipeline that evaluates \texttt{TODDLERS} many thousands of times per object, for example within an MCMC sampler, therefore pays a negligible cost, identical in character to v1; the size of the underlying parameter grid does not affect this lookup.

The second is computing a single model not already on the grid. The shell evolution for one such cloud takes $\approx\!30$\,s on a single core (measured for an \texttt{SB99} Kroupa population at solar metallicity, $n_\mathrm{cl}=160\,\mathrm{cm}^{-3}$, $M_\mathrm{cl}=10^6\,M_\odot$, on an Apple M4 Pro). Its subsequent time-resolved \texttt{Cloudy} post-processing consists of many independent single-timepoint models, which we run in parallel across cores. This cost is paid only when a user wants an off-grid model, never in the table-lookup path above.

The third is generating the grid itself. The same structure scales to the full grid. Each grid point is one dynamics run followed by its time-resolved \texttt{Cloudy} post-processing, and the latter (a large number of single-timepoint models) dominates the cost. The post-processing proceeds in a few ordered passes, because some passes build on the output of an earlier one: while the shell is still embedded in the unswept cloud, the model is constructed from the shell density structure computed in the first pass together with the surrounding unswept cloud, and the optional diffuse-gas model is illuminated by the radiation that leaks out of these inner models. Within any one pass, and across grid points, the individual models are independent. The work is therefore distributed over a worker pool, and the wall-clock time is essentially the total CPU time divided by the number of available cores, independent of the size of the grid. The public release includes the worker-pool pipeline that distributes this work across a cluster allocation.


\end{document}